\documentclass[amsmath,12pt,amssymb,preprint,prd,aps,nofootinbib]{revtex4}
\usepackage{amsfonts} %\usepackage{citesort}
\usepackage{graphicx} % Include figure files
\usepackage{epsfig}
\usepackage{multirow}
\usepackage{bm}% bold math
\begin{document}

\title{\textbf{QCD sum rule analysis of Heavy Quarkonia in magnetized matter - effects of (inverse) magnetic catalysis}}
\author{Pallabi Parui}
\email{pallabiparui123@gmail.com} 
\author{Sourodeep De}
\email{sourodeepde2015@gmail.com}
\author{Ankit Kumar}
\email{ankitchahal17795@gmail.com} 
\author{Amruta Mishra}
\email{amruta@physics.iitd.ac.in}  
\affiliation{Department of Physics, 
Indian Institute of Technology, Delhi, Hauz Khas, New Delhi - 110016}

\begin{abstract}
In-medium masses of the $1S$ and $1P$ states of heavy quarkonia are investigated 
in the magnetized asymmetric nuclear medium, accounting for the Dirac sea effects, using a combined approach of chiral effective model and QCD sum rule method. Masses are calculated from the in-medium scalar and twist-2 gluon condensates, calculated within the chiral model. The gluon condensate is simulated 
through the scalar dilaton field, $\chi$ introduced in the model 
through a scale-invariance breaking logarithmic potential. 
Contribution of the Dirac sea is incorporated through the nucleonic tadpole diagrams. Treating the scalar fields as classical, the dilaton field, $\chi$, the isoscalar non-strange $\sigma (\sim (\langle \bar u u\rangle +\langle \bar d d\rangle ))$, strange $\zeta (\sim \langle \bar s s\rangle)$ and isovector $\delta (\sim (\langle\bar u u\rangle-\langle\bar d d\rangle)$) fields, are obtained by solving their coupled equations 
of motion as derived from the chiral model Lagrangian. The effects of magnetic field are incorporated through the Dirac sea as well as the Landau energy levels of protons, and the anomalous magnetic moments of nucleons. The scalar fields modify appreciably with magnetic field due to the Dirac sea contribution in comparison to the case when it is not considered. In-medium masses of the charmonium and bottomonium
ground states are observed to have significant modifications with magnetic field due to the effects of (inverse) magnetic catalysis. In presence of an external magnetic field, there is mixing between the longitudinal component of the vector and the pseudoscalar mesons (PV mixing) in both quarkonia sectors, leading to a rise (drop) of the masses of $J/\psi^{||}\ (\eta_c$) and $\Upsilon^{||}(1S)\ (\eta_b$) states. These might show in the experimental observables, e.g., the dilepton spectra in the peripheral, ultra-relativistic 
heavy ion collision experiments at RHIC and LHC, where the produced magnetic field is huge.
\end{abstract}
\maketitle
\vspace{-1cm}
\section{Introduction}
The study of the in-medium properties of hadrons is an important area of research in the physics of strongly interacting matter. The study of the heavy flavor hadrons \cite{Hosaka} has attracted a lot of attention due to its relevance in the ultra-relativistic heavy ion collision experiments. Recently, heavy quarkonia ($\bar{q}q; q= c,b$) under extreme conditions of matter i.e., high density and/or high temperature, have been investigated extensively. The medium created in the relativistic, high energy collisions affect the masses and decay widths of the produced particles and have significant observable impacts, e.g., the production and propagation of the particles. In the peripheral heavy ion collisions, strong magnetic fields are expected to be produced, at RHIC in BNL and LHC in CERN \cite{kharzeev,fukushima,skokov,deng, tuchin}. However, the time evolution of the produced magnetic field requires detailed knowledge of the electrical conductivity 
of the medium and proper treatment of the solutions of magneto-hydrodynamic 
equations \cite{tuchin}, which is still an open question. The study of the effects of strong magnetic fields on the in-medium properties of hadrons has 
initiated a new area of research in the physics of heavy ion collisions. The heavy quarkonia are the bound states of a heavy quark ($q=c$ or, $b$) and its antiquark. Charmonium ($\bar{c}c$) and bottomonium ($\bar{b}b$) states have been investigated in the literature using a variety of approaches, i.e., the potential models \cite{eichten1,eichten2,radfort}, the QCD sum rule approach \cite{klingl,kim,am82,temp85,temp897,cpc43,epja79, pallabi, pallabibottom, prd42, song, morita0}, the coupled channel approach \cite{molina}, quark-meson coupling model \cite{krein,tsushima}, a chiral effective model \cite{amupsarx, am981, am90, cpc46}, and a field theoretic model 
for composite hadrons \cite{am102, am95}.

In-medium masses of the ground states of heavy quarkonium in a hadronic medium, have been studied extensively in the literature using the non-perturbative QCD sum rule (QCDSR) approach. In the isospin asymmetric hot nuclear medium, in-medium masses have been studied from the medium modified scalar and twist-2 gluon condensates, calculated within the chiral $SU(3)$ model in terms of the scalar dilaton field $\chi$ and other scalar fields, in the absence of a magnetic field \cite{am82}, and in presence of an external magnetic field \cite{epja79, cpc43}. At finite magnetic field, Landau energy levels of protons and anomalous magnetic moments (AMMs) of the nucleons contribute to the scalar fields through number ($\rho_{p,n}$) and scalar ($\rho^s_{p,n}$) densities of the nucleons within the magnetized nuclear matter \cite{pallabi, pallabibottom, epja79, cpc43}. Thermal modifications of the $S$-waves bottomonium spectral functions have been investigated using QCDSR with the maximum entropy method \cite{temp897}. The temperature effects were included through the gluon condensates, which were estimated from the finite temperature lattice QCD data. Investigation of the effects of finite temperature and baryon chemical potential on the mass of the charmonium states were performed by using the QCD perturbative (second-order stark effect) and non-perturbative sum rule methods \cite{temp85}. The medium effects of temperature and density were incorporated through the gluon condensates calculated in a resonance gas model. In \cite{am102, cho91,cho14}, the magnetically induced mixing between the pseudoscalar and (longitudinal component of) vector charmonium mesons ($\eta_c$-$J/\psi^{||}$) have been investigated using a hadronic effective Lagrangian which leads to a level repulsion between the masses of $J/\psi^{||}$ and $\eta_c$ with increasing magnetic field. In \cite{cho91}, the mass shifts of $\eta_c$ and $J/\psi$ also have been studied using QCDSR framework, considering the mixing effects through the current correlator in the phenomenological side. The OPE side contained the effects of magnetic field in terms of its operator expectation value and the vacuum scalar gluon condensate term up to dimension-4. In the chiral effective model, mass shifts of the heavy quarkonia are obtained through the modifications in the scalar gluon condensates \cite{schechter}, given in terms of the medium modified scalar dilaton field $\chi$, within the chiral SU(3) model \cite{am981, am90}.

The enhancement of the light quark condensates with increasing magnetic field, is called magnetic catalysis \cite{kharzeevmc, kharmc1, elia, chernodub}. In the literature, this effect has been studied to a large extent on the quark matter sector using the Nambu-Jona-Lasinio (NJL) model \cite{Preis, menezes, ammc}. In reference \cite{haber}, magnetic catalysis (MC) have been studied in the context of nuclear matter, through the contributions of magnetized Dirac sea within the Walecka model and an extended linear sigma model. Magnetic catalysis have been observed through the rise in the scalar field $\sigma (\sim \langle \bar{q}q\rangle)$ with magnetic field in the vacuum, for zero AMM of the nucleons. As a consequence, the effective nucleon mass, $m^*_N=m_N-g_{\sigma N}\sigma$, increases with magnetic field in the vacuum ($m_N$ is the vacuum mass of the nucleon and $g_{\sigma N},$ the $\sigma$-nucleon coupling constant in the Lagrangian). In \cite{arghya}, the effects of (inverse) magnetic catalysis have been studied using the weak-field approximation of fermion propagator. The critical temperature of vacuum to nuclear matter phase transition decreases with magnetic field, for the nonzero AMMs of the nucleons, implying an inverse magnetic catalysis (IMC) \cite{balicm}. The zero value for AMM leads to an opposite behavior, namely to the magnetic catalysis. In the literature there are few works related to the effect of IMC/MC on hadronic properties in the nuclear matter.

 The in-medium masses of the open heavy flavor mesons, namely the open charm and open bottom mesons, have been studied within the QCD sum rule approach \cite{arata,wang, gubler}, and using the generalized version of the chiral effective model, both in the absence of an external magnetic field \cite{am23, amd91}, and in presence of a magnetic field \cite{am97, am98}. The open heavy flavor mesons have their mass modifications in terms of both the light quark condensates (because of the light quark flavor present in their quark structure) as well as the gluon condensates simulated within the chiral model by the scalar dilaton field $\chi$. The in-medium masses of the light vector mesons ($\rho,\omega,\phi$) have been studied within the QCD sum rule approach \cite{hatsuda}. The medium modifications of the masses obtained from the non-strange ($\langle \bar{q}q\rangle; q=u,d$ for $\rho,\omega$) and strange ($\langle \bar{s}s\rangle$ for $\phi$) light quark condensates and the scalar gluon condensates ($\sim\langle G^a_{\mu\nu}G^{a\mu\nu}\rangle$), calculated within the chiral $SU(3)$ model, in the strange asymmetric matter in absence of magnetic field \cite{am91}, and in the magnetized nuclear medium \cite{am100}. 
 
 In the magnetized nuclear matter, the mass modifications of the $1S \ (J/\psi)$, $2S\ (\psi(3686))$ and $1D\ (\psi(3770))$ states of charmonium have been studied in terms of the medium modifications of the scalar dilaton field $\chi$, which in turn mimics the gluon condensates of QCD within the chiral effective model \cite{am981}. For the charmonium states, there observed to be a mass drop as the density increases beyond the nuclear matter saturation density $\rho_0$, for different values of magnetic field, $|eB|$ and isospin asymmetry parameter, $\eta$. The dominant effect was coming from the nuclear matter density as compared to the effects from the magnetic field. In \cite{ko}, the mass shifts of these charmonium states due to the change in the gluon condensate have been calculated using the perturbative QCD approach, to the leading order in density till $\rho_0$. The mass shifts obtained in \cite{am981}, calculated for any baryonic density within the chiral effective model, agrees with the results of \cite{ko} in the linear density approximation of $-8$, $-100$ and $-140$ (in MeV) for $J/\psi$, $\psi(3686)$ and $\psi(3770)$, respectively at $\rho_B=\rho_0$ and $|eB|=0$. In-medium masses of the open charm and charmonium mesons have also been studied in the magnetized strange hadronic matter \cite{cpc46}. 
The PV mixing between the longitudinal component of vector and pseudoscalar open charm ($D^{*||}-D$) mesons \cite{amarx} and $S$-waves of charmonia \cite{am102,amarx} have been studied using a hadronic effective Lagrangian. The in-medium masses thus obtained have been used to study the in-medium hadronic decay widths for $D^*\rightarrow D\pi$ \cite{amarx} and of $\psi(3770)\rightarrow D\bar{D}$ \cite{am102,amarx}, accounting for the lowest Landau energy level contributions for the charged $D$ mesons. In \cite{amsm1}, the spin-magnetic field interaction between $B-B^*$ and $\eta_b(4S)-\Upsilon(4S)$ have been studied using a Hamiltonian approach \cite{alford} in the presence of an external magnetic field. The in-medium partial decay widths of $\Upsilon(4S)$ going to $B\bar{B}$ have been studied using a field theoretic model for composite hadrons with quark (and antiquark) constituents \cite{ amsm1}. The in-medium decay widths of charmonium states to $D\bar{D}$ in the magnetized nuclear matter have also been studied using a light quark-antiquark pair creation model, namely the $^3P_0$ model \cite{am3p0}. The spin-magnetic field interaction between the spin-singlet and longitudinal component of the spin-triplet states have been studied using a Hamiltonian formalism of \cite{alford} on the $1S$ states of heavy quarkonia ($\eta_c-J/\psi$) and ($\eta_b-\Upsilon(1S)$), which lead to a rise (drop) in the mass of the $J/\psi^{||}\ (\eta_c)$ and $\Upsilon(1S)^{||}\ (\eta_b)$ with increasing magnetic field \cite{alford, pallabi, pallabibottom}. The effects of (inverse) magnetic catalysis due to the Dirac sea contribution at finite $|eB|$, have not been considered in the studies mentioned above. The studies of the magnetic field modified hadronic properties, specifically of the heavy flavor mesons have important observable consequences, such as in the formation time of heavy quarkonia, the particle production ratio, etc. in the non-central ultra relativistic heavy ion collision experiments \cite{machado, suzuki}. 

In the present work, we study the in-medium masses of the $1S$-wave (vector,
$J/\psi$, pseudoscalar, $\eta_c$) and $1P$-wave (scalar, $\chi_{c0}$, axial-vector, $\chi_{c1}$) charmonium states as well as the $1S$-wave (vector, $\Upsilon (1S)$, pseudoscalar, $\eta_b$) and $1P$-wave (scalar, $\chi_{b0}$, axial-vector, $\chi_{b1}$) bottomonium states, in a magnetized isospin asymmetric nuclear medium, using the QCD sum rule approach, by incorporating the effects of Dirac sea in presence of an external magnetic field. At finite magnetic field, effects of the pseudoscalar-vector (PV) mixing between the pseudoscalar $\eta_c$ ($\eta_b$) and longitudinal component of vector $J/\psi^{||}$ ($\Upsilon^{||}(1S)$) mesons are studied in both quarkonia sector, accounting for the Dirac sea effects on their in-medium masses. 

The outline of the paper is: in section II, the chiral effective model is described briefly to calculate the medium modified gluon condensates. Section III illustrates the QCD Sum Rule framework to calculate the in-medium masses of the lowest lying states of heavy quarkonia. Mass shifts of the S-wave states due to the pseudoscalar-vector (PV) mesons mixing are introduced in presence of a magnetic field. In section IV, results of the magnetized Dirac sea effects are discussed. Section V summarizes the findings of the present work.

\section{The Chiral ${SU(3)}_L\times{SU(3)}_R $ Model}

In-medium masses of the quarkonium ground states are computed within the QCD sum rule approach, in terms of the scalar and twist-2 gluon condensates. In the present study, these condensates are calculated within an effective chiral hadronic model \cite{papa59}. The chiral model is based on the non-linear realization of chiral ${SU(3)}_L\times{SU(3)}_R$ symmetry \cite{weinberg, coleman, bardeen}, and the broken scale invariance of QCD \cite{papa59, am69, zschi}. The QCD scale-invariance breaking is incorporated through a logarithmic potential in the scalar dilaton field $\chi$ \cite{erik}, in the model. The chiral  model Lagrangian density has the following general form \cite{papa59},
\begin{equation}
   \mathcal{L}=\mathcal{L}_{kin}+\mathcal{L}_{BM}+\mathcal{L}_{vec}+\mathcal{L}_0+\mathcal{L}_{scale-break}+\mathcal{L}_{SB}+\mathcal{L}_{mag} 
\end{equation}
in the above expression, $\mathcal{L}_{kin}$ is the kinetic energy of the baryons and the mesons; $\mathcal{L}_{BM}$ represents the baryon-mesons (both spin-0 and spin-1 mesons) interactions; $ \mathcal{L}_{vec}$ contains the quartic self-interactions of the vector mesons and their couplings with the scalar ones; $\mathcal{L}_0$ incorporates the spontaneous chiral symmetry breaking effects via meson-meson interactions; $\mathcal{L}_{scale-break}$ is the QCD scale symmetry breaking logarithmic potential;  the explicit symmetry breaking term is $\mathcal{L}_{SB}$; finally the magnetic field effects on the charged and neutral baryons in the nuclear medium are given by \cite{amupsarx, am981, am97, am98, broderik, prakash, wei, guang},
\begin{equation}
\mathcal{L}_{mag}=-\frac{1}{4}F_{\mu\nu}F^{\mu\nu}-q_i{\bar{\psi}}_i\gamma_\mu A^\mu\psi_i-\frac{1}{4}\kappa_i\mu_N{\bar{\psi}}_i\sigma^{\mu\nu}F_{\mu\nu}\psi_i
\end{equation}
where,  $\psi_i$ is the baryon field operator ($ i = p, n$, in case of nuclear matter), the parameter, $\kappa_i$ is related to the anomalous magnetic moment of the i-th baryon, $\kappa_p = 3.5856$ and $\kappa_n = -3.8263$, are the gyromagnetic ratio corresponding to the anomalous magnetic moments (AMM) of the proton and the neutron respectively \cite{broderik, prakash, wei, guang, ivanov, paoli, dexheimer,franzon}. In the magnetized nuclear medium there are contributions from the protons Landau energy levels \cite{ivanov}, and the nucleons anomalous magnetic moments \cite{ivanov, paoli}, to the number and scalar densities ($\rho_i,\ \rho^s_i,i=p,n$, respectively) of the nucleons \cite{am97, am98}. In the current study, the Dirac sea contributes to the scalar densities of nucleons at finite magnetic field, including the effects of the anomalous magnetic moments of nucleons within the chiral $SU(3)$ model. One-loop self energy functions of the nucleons are evaluated through summation over the scalars ($\sigma$, $\zeta$ and $\delta$) and vectors ($\rho$ and $\omega$) tadpole diagrams, using the weak-field expansion of the nucleonic propagator \cite{arghya}, accounting for the AMMs of nucleons, within the chiral effective model.  

The meson fields of the chiral model Lagrangian are treated as the classical fields, whereas the nucleons as the quantum fields in the evaluation of the Dirac sea contribution to the scalar fields. The scalar dilaton field, $\chi$  simulates the scalar gluon condensate $ \langle \frac{\alpha_s}{\pi} G_{\mu\nu}^a$ $G^{a\mu\nu}\rangle$, as well as the twist-2 gluon operator $ \langle \frac{\alpha_s}{\pi} G_{\mu\sigma}^a$ $G_{\nu}^{a\space \sigma}\rangle$, within the model. The energy momentum tensor, $T_{\mu\nu}$ derived from the $\chi$-dependent terms in the chiral model Lagrangian density is thus \cite{am82}  
\begin{equation}
 T_{\mu\nu}=\left(\partial_{\mu}\chi\right)\left(\frac{\partial\mathcal{L}}{\partial\left(\partial^\nu\chi\right)}\right)- g_{\mu\nu}\mathcal{L}_\chi
\end{equation}      
The QCD energy momentum tensor, in the limit of finite current quark mass contains a symmetric trace-less part and a trace part \cite{morita}, 
\begin{equation}
T_{\mu\nu}=-ST\left(G_{\mu\sigma}^a G_\nu^{a\sigma}\right)+\frac{g_{\mu\nu}}{4}\left(\sum_{i}m_i\bar{q}_i q_i+ \frac{\beta_{QCD}}{2g}(G_{\sigma k}^a G^{a\sigma}_k)\right) 
\end{equation}
with the leading order QCD $\beta$ function \cite{am82}, $\beta_{QCD}(g) = -\frac{g^3}{(4\pi)^2} (11-\frac{2}{3} N_f )$, for three color quantum numbers of QCD, and $N_f=3$ number of quark flavors. Here, $m_i$'s $(i= u, d, s)$ are the current quark masses. The medium expectation value of the twist-2 gluon operator is, 
\begin{equation}
     \langle \frac{\alpha_s}{\pi} G_{\mu\sigma}^a G_{\nu}^{a\space \sigma}\rangle = \left(u_\mu u_\nu-\frac{g_{\mu\nu}}{4}\right)G_2
\end{equation}
where $u_\mu$ is the 4-velocity of the nuclear medium taken to be at rest in the present investigation, namely, $u_\mu=\left(1,\ 0,\ 0,\ 0\right)$. The energy momentum tensor of QCD is thus \cite{am82}
\begin{equation}
    T_{\mu\nu}=-\frac{\pi}{\alpha_s}\left(u_\mu u_\nu-\frac{g_{\mu\nu}}{4}\right)G_2+\frac{g_{\mu\nu}}{4}\left(\sum_{i}m_i\langle\bar{q}_i q_i\rangle+ \frac{\beta_{QCD}}{2g}\langle G_{\sigma k}^a G^{a\sigma}_k\rangle\right)
\end{equation}
Comparing the expressions of energy momentum tensor from  equations (6) and (3), the expressions for $G_2$ (the twist-2 component) and the scalar gluon condensate are given by multiplying both sides with $\left(u_\mu u_\nu-\frac{g_{\mu\nu}}{4}\right)$ and $g^{\mu\nu}$ respectively. These are given by
\begin{equation}
G_2 = \frac{\alpha_s}{\pi}\Bigg[-(1-d+4k_4)(\chi^4-\chi_0^4)-
\chi^4 \ln\left(\frac{\chi^4}{\chi_0^4}\right)+\frac{4}{3}d\chi^4\ln\left(\left(\frac{(\sigma^2-\delta^2)\zeta}{\sigma_0^2\zeta_0}\right)\left(\frac{\chi}{\chi_0}\right)^3\right)\Bigg]
\label{g2}
\end{equation}
and \cite{cohen}, 
\begin{equation}
 \langle \frac{\alpha_s}{\pi} G_{\mu\nu}^a G^{a\mu\nu}\rangle
=\frac{8}{9}\Bigg[(1-d)\chi^4+\left(\frac{\chi}{\chi_0}\right)^2\left(m_\pi^2 f_\pi \sigma +\left(\sqrt{2}m_k^2f_k-\frac{1}{\sqrt{2}}m_\pi^2 f_\pi\right)\zeta\right)\Bigg]
\label{g1}
\end{equation}
The expectation values of the scalar and the twist-2 gluon condensates in magnetized nuclear medium, depend on the in-medium values of the non-strange scalar-isoscalar field $\sigma$, the strange scalar-isoscalar field $\zeta$, the non-strange scalar-isovector field $\delta$ and the scalar dilaton field $\chi$, within the chiral $ SU(3) $ model. The Euler Lagrange's equations of motion for the scalar fields are derived from the chiral model Lagrangian. Contributions of magnetic fields to the scalar fields are obtained through the scalar ($\rho^s_{p,n}$) and number ($\rho_{p,n}$) densities of the nucleons, which are modified due to the Landau energy levels of protons and the anomalous magnetic moments of nucleons. Effects of the magnetized Dirac sea also contributes to the scalar densities of nucleons and hence on the scalar fields, at zero and finite density matter.

\section{In-Medium Masses Within The QCD Sum Rule Approach}
The in-medium masses of the $1S$-wave and $1P$-wave states of charmonium [$1S$: $J/\psi$, $\eta_c$ and $1P$: $\chi_{c0}$, $\chi_{c1}$] and bottomonium [$1S$: $\Upsilon (1S)$, $\eta_b$ and $1P$: $\chi_{b0}$, $\chi_{b1}$] are studied within the QCD Sum Rule approach. In this approach, masses of the heavy-quarkonium ground states are calculated in terms of the medium modified scalar and twist-2 gluon condensates. The condensates are obtained within the chiral model, through the in-medium values of the scalar fields in the magnetized, asymmetric nuclear matter with the additional contribution from the magnetized Dirac sea. The in-medium mass squared, $m_i^{*2}$ of the i-type of quarkonium ground state (i= vector, pseudoscalar, scalar, and axial-vector) is given as \cite{reinders}, 
\begin{equation}
    m_i^{*2}\simeq \frac{M_{n-1}^i (\xi)}{M_{n}^i (\xi)}-4m_q^2\xi
\end{equation}
    Where $M_{n}^i$ is the $n$-th moment of the i-type meson state and $m_q\ (q=c,b)$ is the running heavy quark mass dependent on the renormalization scale $\xi$.
    Using the operator product expansion technique [OPE], the moment can be written as \cite{klingl, reinders}, 
    \begin{equation}
      M_{n}^i(\xi)=A_n^i(\xi)\left[1+ a_n^i(\xi)\alpha_s + b_n^i(\xi)\phi_b+c_n^i(\xi)\phi_c\right] 
    \end{equation}
Where $A_n^i, a_n^i, b_n^i,$ and $c_n^i$ are the Wilson coefficients. The coefficients, $A_n^i$ result from the bare-loop diagram of perturbative QCD, $ a_n^i$ are the contributions of the perturbative radiative corrections, and $b_n^i$ are related to the scalar gluon condensate through 
\begin{equation}
    \phi_b=\frac{4\pi^2}{9}\frac{\langle \frac{\alpha_s}{\pi} G_{\mu\nu}^a G^{a\mu\nu}\rangle}{\left(4m_q^2\right)^2}.
\end{equation}
By substituting the expression for the scalar gluon condensate from equation (8),
\begin{equation}
    \phi_b=\frac{32\pi^2}{81\left(4m_q^2\right)^2}\Bigg[(1-d)\chi^4+ \left(\frac{\chi}{\chi_0}\right)^2\left(m_\pi^2 f_\pi \sigma +\left(\sqrt{2}m_k^2f_k-\frac{1}{\sqrt{2}}m_\pi^2 f_\pi\right)\zeta\right)\Bigg]
\label{phib}
\end{equation}
The coefficients $c_n^i$ are associated with the twist-2 gluon condensates as 
\begin{equation}
    \phi_c=\frac{4\pi^2}{3\left(4m_q^2\right)^2 }G_2
\end{equation}
which using equation (7), reduces to 
\begin{multline}
 \phi_c =  \frac{4\pi \alpha_s}{3\left(4m_q^2\right)^2} 
\Bigg[-(1-d+4k_4)(\chi^4-\chi_0^4)-
\chi^4 \ln\left(\frac{\chi^4}{\chi_0^4}\right)  
+\frac{4}{3}d\chi^4\ln\left(\frac{(\sigma^2-\delta^2) \zeta  \chi^3}{\sigma_0^2\zeta_0 \chi^3_0}\right)\Bigg]
\label{phic}   
\end{multline}
%\begin{eqnarray}
  %  \phi_c &= & \frac{4\pi \alpha_s}{3\left(4m_q^2\right)^2} 
%\Bigg[-(1-d+4k_4)(\chi^4-\chi_0^4)-
%\chi^4 \ln\left(\frac{\chi^4}{\chi_0^4}\right) \nonumber \\ 
%&+&\frac{4}{3}d\chi^4\ln\left(\left(\frac{(\sigma^2-\delta^2)\zeta}{\sigma_0^2\zeta_0}\right)\left(\frac{\chi}{\chi_0}\right)^3\right)\Bigg]
%\label{phic}
%\end{eqnarray}
The $\xi$-dependent parameters $m_q$ ($q=c,b$) and the running coupling constant $\alpha_s$ are \cite{am82, reinders}
\begin{equation}
    \frac{m_q(\xi)}{m_q} = 1-\frac{\alpha_s}{\pi}\left[\frac{2+\xi}{1+\xi} \ln(2+\xi)-2\ln2\right]
\end{equation}
with $m_c\equiv m_c (p^2=-m_c^2)=1.26 $ GeV and  $m_b\equiv m_b (p^2=-m_b^2)=4.23 $ GeV \cite{am82, reinders85}, and 
\begin{equation}
    \alpha_s \left(Q_0^2+4m_q^2\right) = \alpha_s(4m_q^2) \Bigg/ \left(1+\frac{(33-2n_f)}{12\pi}\alpha_s(4m_q^2)\ln\frac{Q_0^2+4m_q^2}{4m_q^2}\right)
\end{equation}
with $Q_0^2=4m_q^2 \xi$ $(q=c,b)$, and $n_f=4$, $\alpha_s \left(4m_c^2\right) \simeq 0.23$ in the charm quark sector, and $n_f=5$, $\alpha_s \left(4m_b^2\right) \simeq 0.15$  in the bottom quark sector \cite{reinders85}. \\
The Wilson coefficients, $A_n^i, a_n^i$ and $b_n^i$ are given in \cite{reinders} for different $J^{PC}$ quantum numbers of states, for e.g., the pseudoscalar, vector, scalar, axial-vector channels. The $c_n^i$s' are listed for the vector and pseudoscalar ($1S$ states) channels in \cite{klingl}, for the $1P$-wave states (scalar and axial-vector) $c_n^i$s' are calculated using a background field technique \cite{song}.\\
At finite magnetic field, mixing of the pseudoscalar ($P\equiv \eta_c(1S)$) 
and vector ($V\equiv J/\psi$) charmonium states
are considered through the interaction Lagrangian \cite{cho91, am102, amarx, gubler, cho14}
\begin{equation}
{\cal L}_{PV\gamma}=\frac{g_{PV}}{m_{av}} e {\tilde F}_{\mu \nu}
(\partial ^\mu P) V^\nu,
\label{PVgamma}
\end{equation}
where $m_{av}=(m_V+m_P)/2$, $m_P$ and $m_V$ are the masses 
for the pseudoscalar and vector charmonium states,
${\tilde F}_{\mu \nu}$ is the dual electromagnetic field strength tensor. In equation (\ref{PVgamma}), the coupling parameter $g_{PV}$
is fitted from the observed value of the radiative decay width, 
\begin{equation}
\Gamma (V\rightarrow P \gamma)
=\frac{e^2}{12}\frac{g_{PV}^2 {p^3_{cm}}}{\pi m_{av}^2},
\label{decay_VP}
\end{equation}
where, $p_{cm}=(m_V^2-m_P^2)/(2m_V)$
is the magnitude of the center of mass
momentum in the final state. 
The masses of the pseudoscalar and the longitudinal component
of the vector mesons including the mixing effects 
are given by
\begin{equation}
m^{2\ {(PV)}}_{P,V^{||}}=\frac{1}{2} \Bigg ( M_+^2 
+\frac{c_{PV}^2}{m_{av}^2} \mp 
\sqrt {M_-^4+\frac{2c_{PV}^2 M_+^2}{m_{av}^2} 
+\frac{c_{PV}^4}{m_{av}^4}} \Bigg).
\label{mpv_long}
\end{equation}
Where $M_+^2=m_P^2+m_V^2$, $M_-^2=m_V^2-m_P^2$ and 
$c_{PV}= g_{PV} |eB|$. 
%Considering the terms in equation (\ref{mpv_long}) up to the second order in $c_{PV}$ and leading order in $(m_V-m_P)/2m_{av}$, we get
%\begin{equation}
 % m^{2\ (PV)}_{P,V^{||}}= m_{P/V}^2 \mp \frac{c_{PV}^2}{M_{-}^2}
 % \label{pvmix}
%\end{equation}
 The effective Lagrangian given by equation 
(\ref{PVgamma}) is observed to lead to the
mass modifications of the longitudinal component of $J/\psi$ and
$\eta_c$ states in the presence of magnetic field. In equation (\ref{mpv_long}), the effects of the magnetized Dirac sea, the Landau quantization of protons and the AMMs of the nucleons are incorporated through the in-medium values of $m_{P,V}$. The in-medium masses of $m_{P}$, $m_{V}$ are calculated using the QCDSR approach.
Effects of the spin-magnetic field interaction have been studied for the $S$-wave states of heavy quarkonia \cite{pallabi, pallabibottom, alford, amsm1, iwasaki}. This leads to a level repulsion between the mass eigenstates of the spin-0 and longitudinal component of the spin-1 states. The interaction leads to a mixing effect between ($\Upsilon(1S)-\eta_b$). Thus, the effective masses of $\Upsilon^{||}(1S)$ and $\eta_b$, accounting for the mass shifts due to the spin-magnetic interaction Hamiltonian (\textbf{$-\vec{\mu}.\vec{B}$}) \cite{alford},  
\begin{equation}
    m^{eff}_{\Upsilon (1S)}= m^*_{\Upsilon(1S)} + \Delta m_{sB},\;\;\;\; 
m^{eff}_{\eta_b}= m^*_{\eta_b} - \Delta m_{sB} 
\label{hm}
\end{equation}
In the above equation, $m^*_{\Upsilon(1S)/\eta_b}$ denotes the in-medium masses of the $1S$-wave bottomonium states calculated within QCDSR, accounting for the Dirac sea effects. $\Delta m_{sB}$ is the mass shift due to the spin-magnetic field interaction, given by
\begin{equation}
    \Delta m_{sB} = \frac{\Delta M}{2}\left((1+{\chi_{sB}}^2)^{1/2}-1\right),\\ \chi_{sB} = \frac{2g\mu_b B}{\Delta M}
\end{equation}
where, $\mu_b = (\frac{1}{3}e)/(2m_b)$ is the bottom quark Bohr magneton with the constituent bottom quark mass, $m_b = 4.7 $ GeV \cite{alford}, $\Delta M = m^*_{\Upsilon(1S)} - m^*_{\eta_b}$, and g is chosen to be 2 (ignoring the anomalous magnetic moments of the bottom quark (anti-quark)). The Hamiltonian approach is taken to study the spin-mixing effects in the bottomonium sector (unlike $\bar{c}c$), due to the lack of experimental data on the bottomonium radiative decay width, $\Upsilon(1S)\rightarrow \eta_b\gamma$. 

\section{Results and Discussions}

\subsection{Charmonium states}
\label{r1}
\begin{figure}[h!]
    \includegraphics[width=1.0\textwidth]{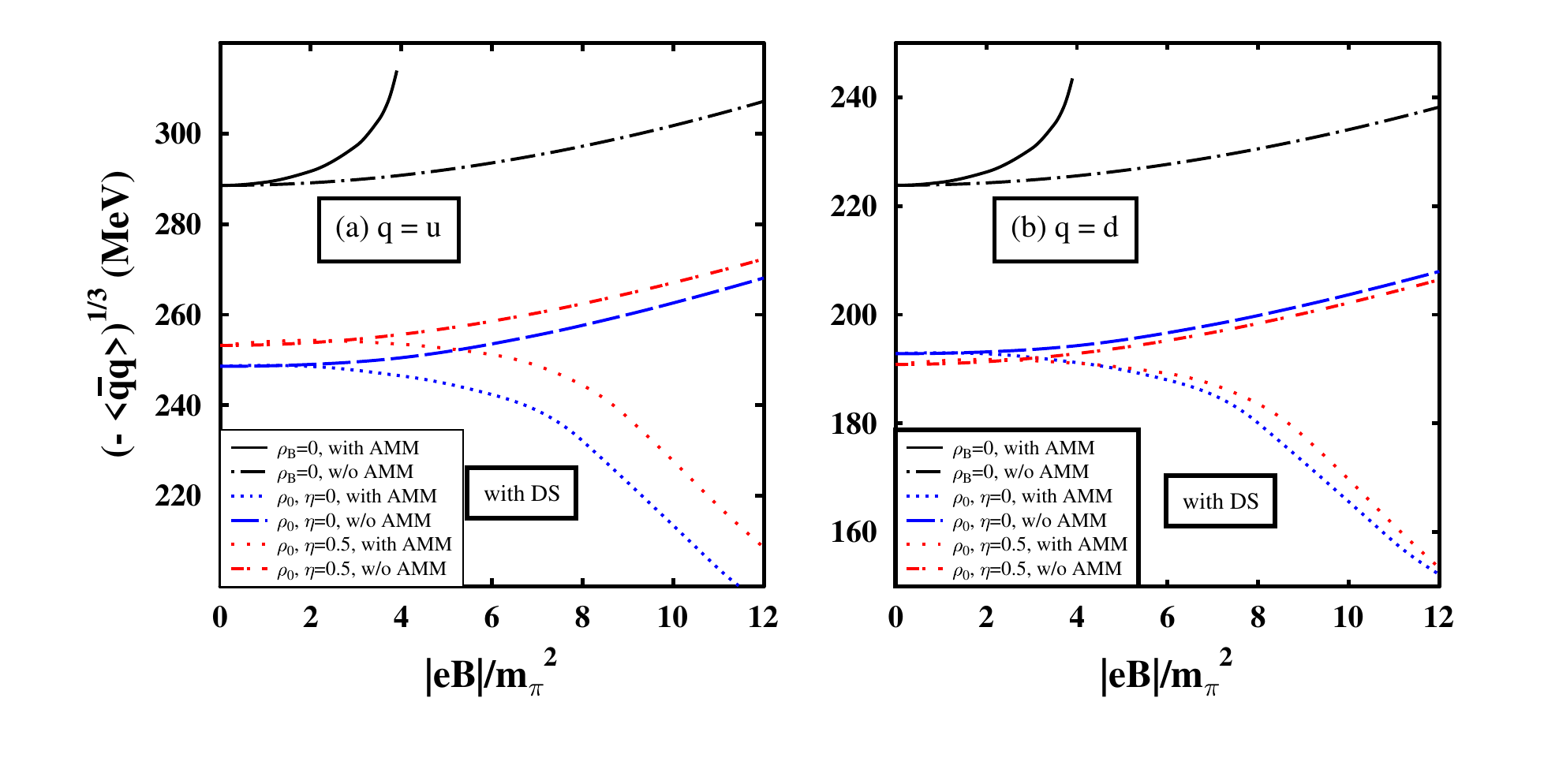}\hfill
    \vspace{-0.8cm}
    \caption{(Color online) The light quark condensates $(-\langle\bar{q}q\rangle)^{1/3}$ in MeV for (a) $q=u$  and (b) $q=d$, are shown as functions of $|eB|/m_{\pi}^2$, at $\rho_B=0$, and $\rho_0$ for $\eta=0,\ 0.5$. The condensates are obtained from the in-medium scalar fields, accounting for the Dirac sea effects. The (drop) rise of the quantity with magnetic field indicate the phenomena of (inverse) magnetic catalysis. The effects of the nucleons' AMMs are considered and compared to the no AMM condition.}
    \label{fig:0}
\end{figure}
\begin{figure}[h!]
    \includegraphics[width=1.0\textwidth]{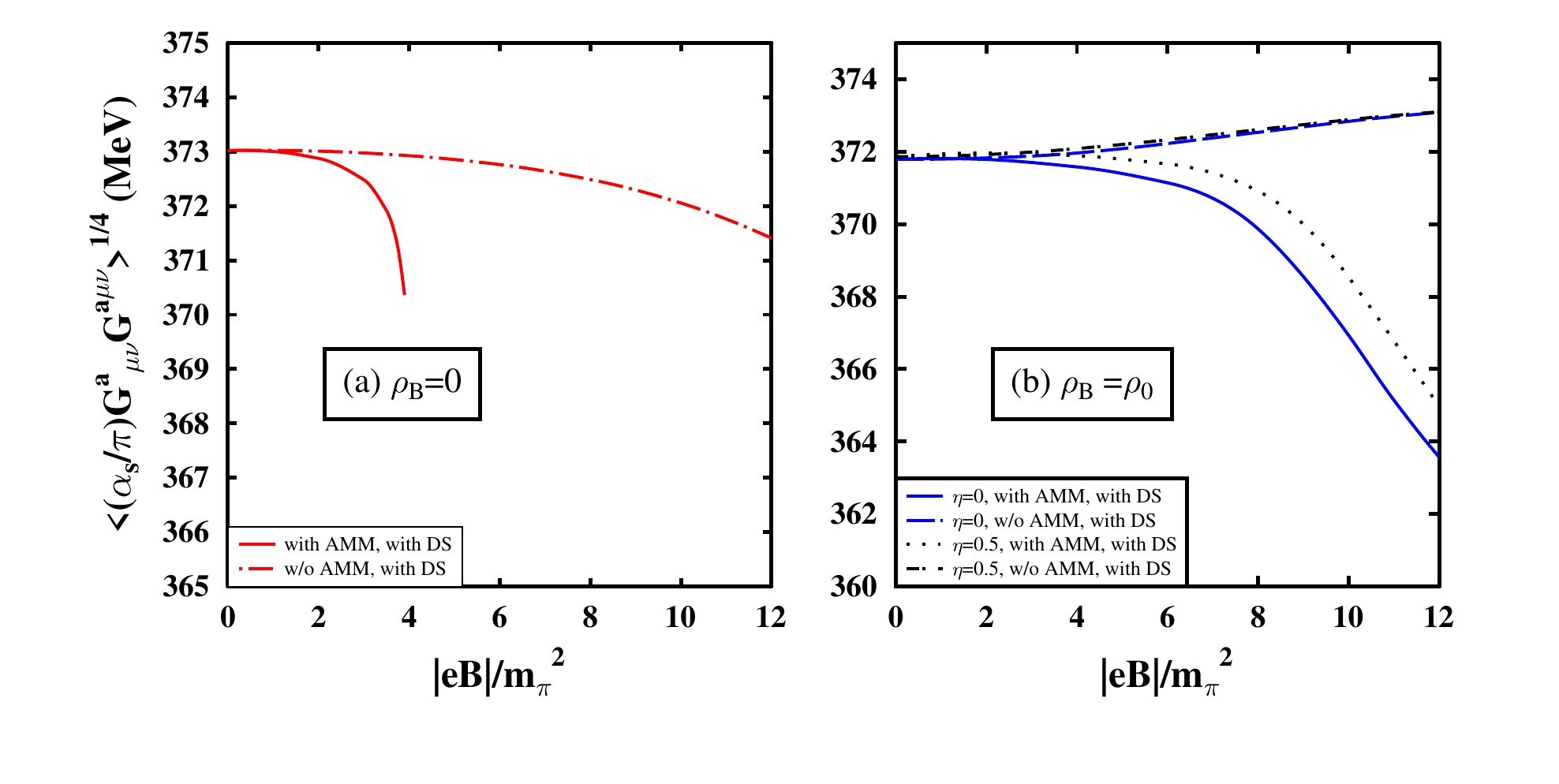}\hfill
    \vspace{-0.8cm}
    \caption{(Color online) The scalar gluon condensate $\langle (\alpha_s/\pi)G^{a}_{\mu\nu}G^{a\mu\nu}\rangle^{1/4}$ in MeV at (a) $\rho_B=0$ and (b) $\rho_B=\rho_0$ (for $\eta=0,\ 0.5$), are shown as functions of $|eB|$ (in units of $m_{\pi}^2$). The condensates are calculated in terms of the in-medium scalar fields, accounting for the Dirac sea (DS) effects. The effects of the nucleons' AMMs are considered and compared with the no AMM case.}
    \label{fig:1}
\end{figure}
\begin{figure}[h!]
    \includegraphics[width=1.0\textwidth]{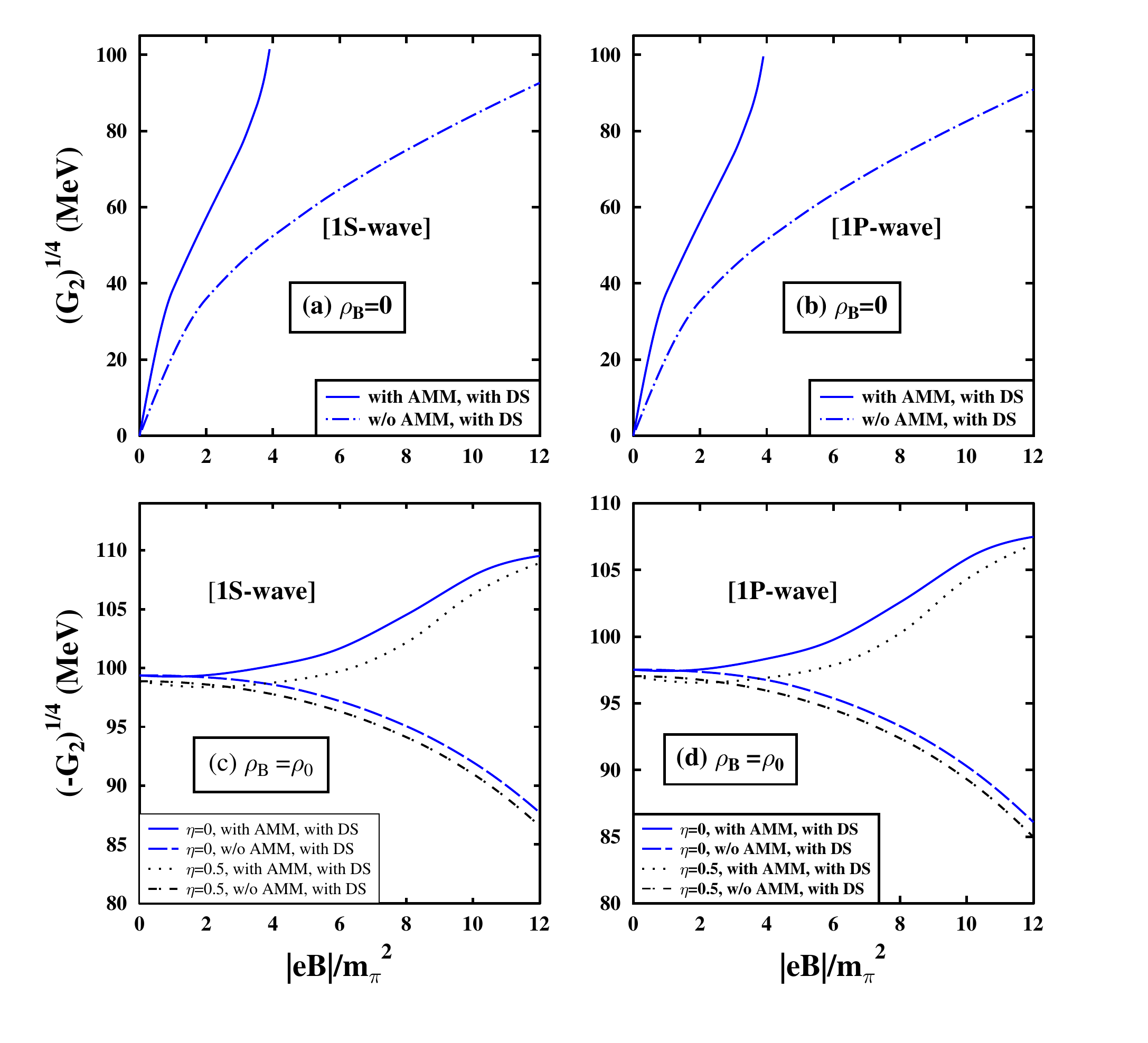}\hfill
    \vspace{-0.8cm}
    \caption{(Color online) The twist-2 gluon condensate $(G_2)^{1/4}$ in MeV are plotted as functions of $|eB|/m_{\pi}^2$, at $\rho_B=0$ [in (a), (b)] and at $\rho_B=\rho_0$ [in (c), (d)] (with appropriate sign as shown). The $G_2$ condensates for the $1S$-wave states in (a) and (c) incorporate the Dirac sea effects in terms of the in-medium scalar fields. Similar plots in (b) and (d) are for the $1P$-wave states also include DS contribution. The magnitudes are slightly different for the $S$ and $P$ waves of charmonium, due to the different values of the running coupling constant $\alpha_s(\xi)$, taking $\xi=1$ for $S$ and $\xi=2.5$ for $P$ waves. Effects of nucleons' AMMs are considered and compared to the no AMM case.}
    \label{fig:2}
\end{figure}
\begin{figure}
    \includegraphics[width=1.0\textwidth]{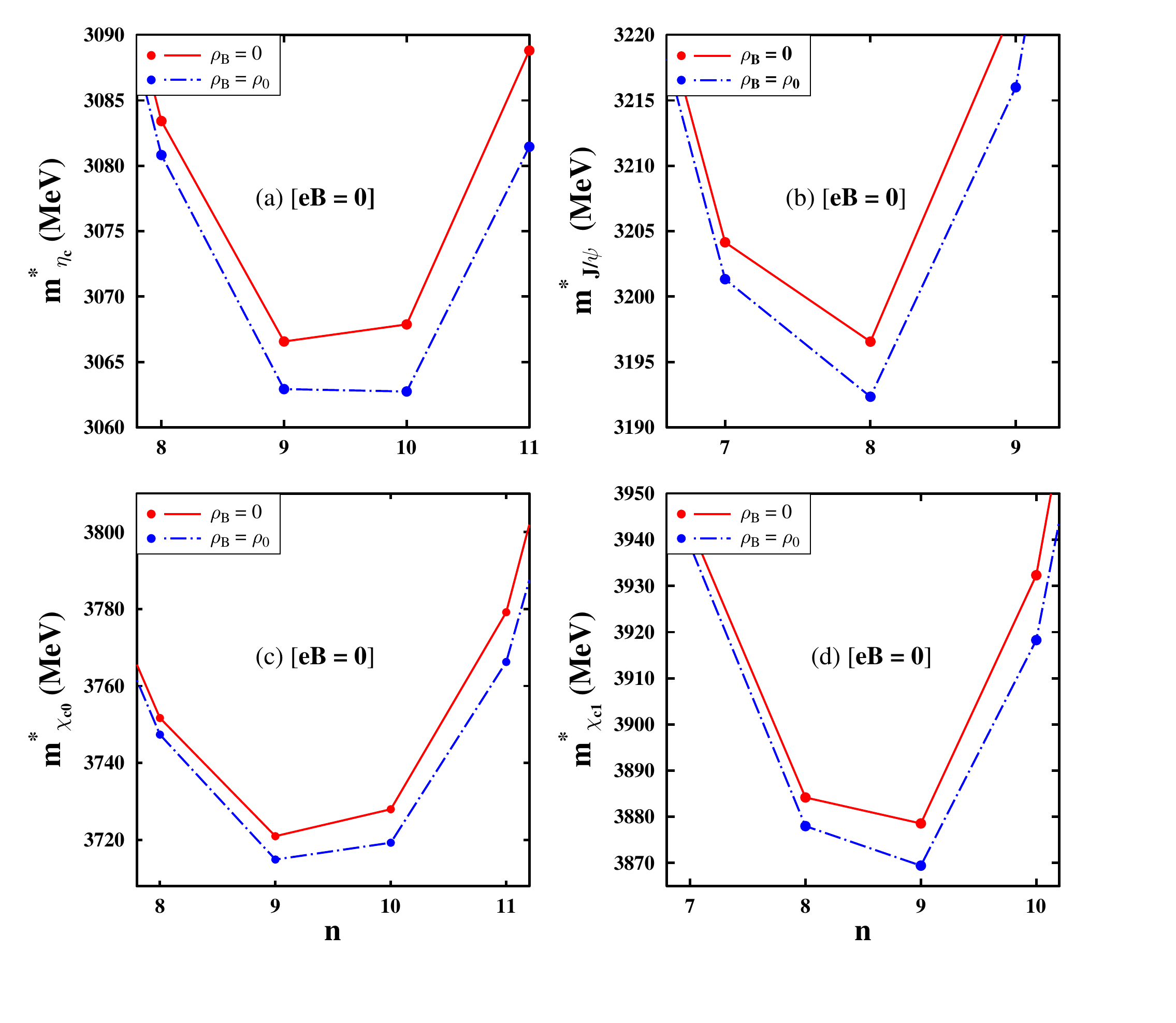}\hfill
    \vspace{-0.8cm}
    \caption{(Color online) $m^*$ (in MeV) are plotted as functions of n, for (a) $\eta_c$, (b) $J/\psi$, (c) $\chi_{c0}$ and (d) $\chi_{c1}$ states of charmonium, at $\rho_B = 0,\ \rho_0$ for symmetric nuclear matter ($\eta=0$), at zero magnetic field $|eB|=0$. The minimum value of $m^{*}$ as a function of n corresponds to the physical mass of that particular state.}
    \label{fig:3}
\end{figure}

\begin{figure}
    \includegraphics[width=1.0\textwidth]{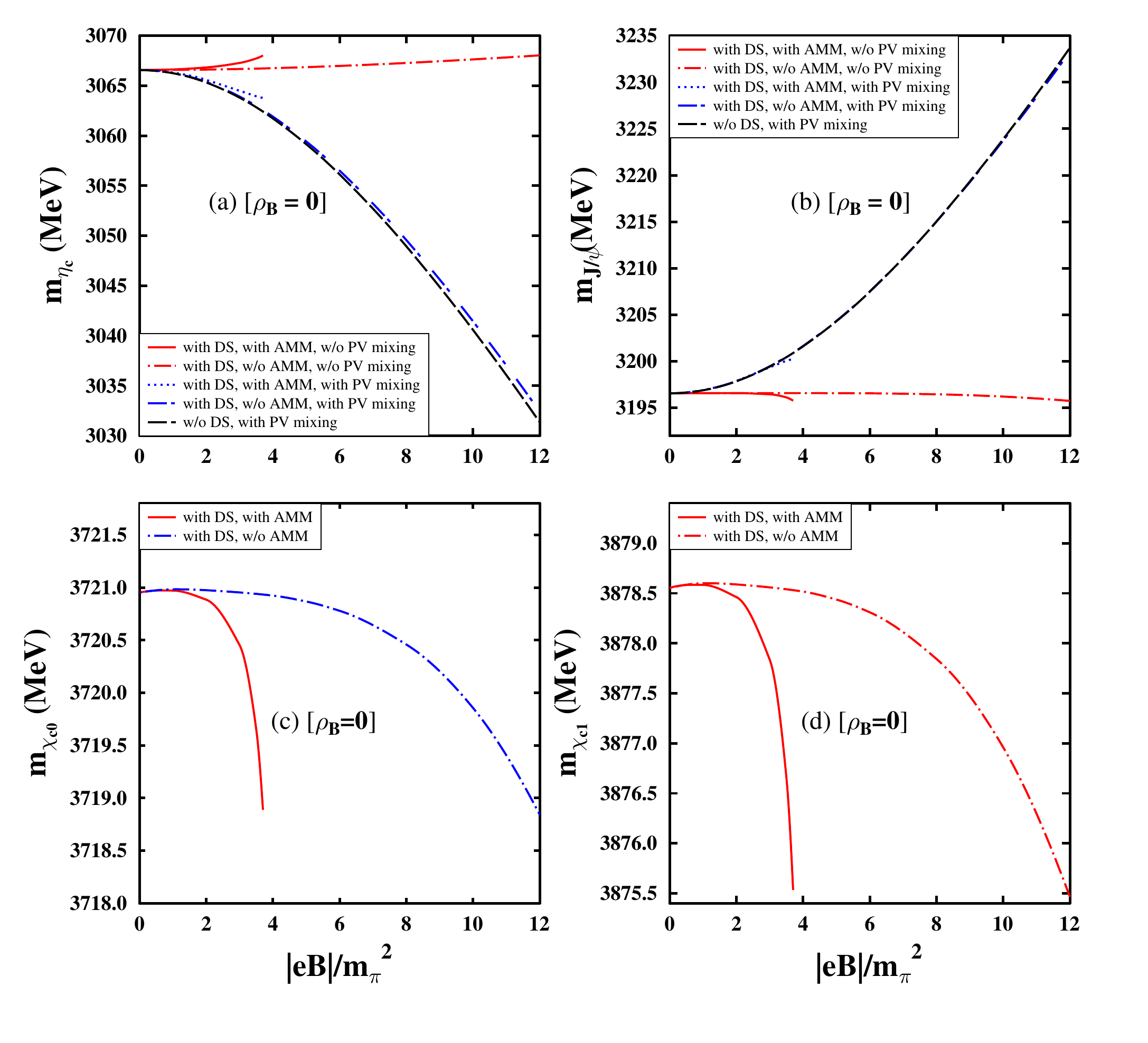}\hfill
    \vspace{-0.8cm}
    \caption{(Color online) Masses (in MeV) are plotted as functions of $|eB|/m_{\pi}^2$, for (a) $\eta_c$, (b) $J/\psi$, (c) $\chi_{c0}$ and (d) $\chi_{c1}$ at $\rho_B = 0$. The effects of the magnetized Dirac sea are shown in the masses, accounting for (and not) the anomalous magnetic moments of the Dirac sea of nucleons. The PV mixing effects between $(J/\psi^{||}-\eta_c)$ are considered, taking into account the DS effects and compared to the case when it is not included.}
    \label{fig:4}
\end{figure}
\begin{figure}
    \includegraphics[width=1.0\textwidth]{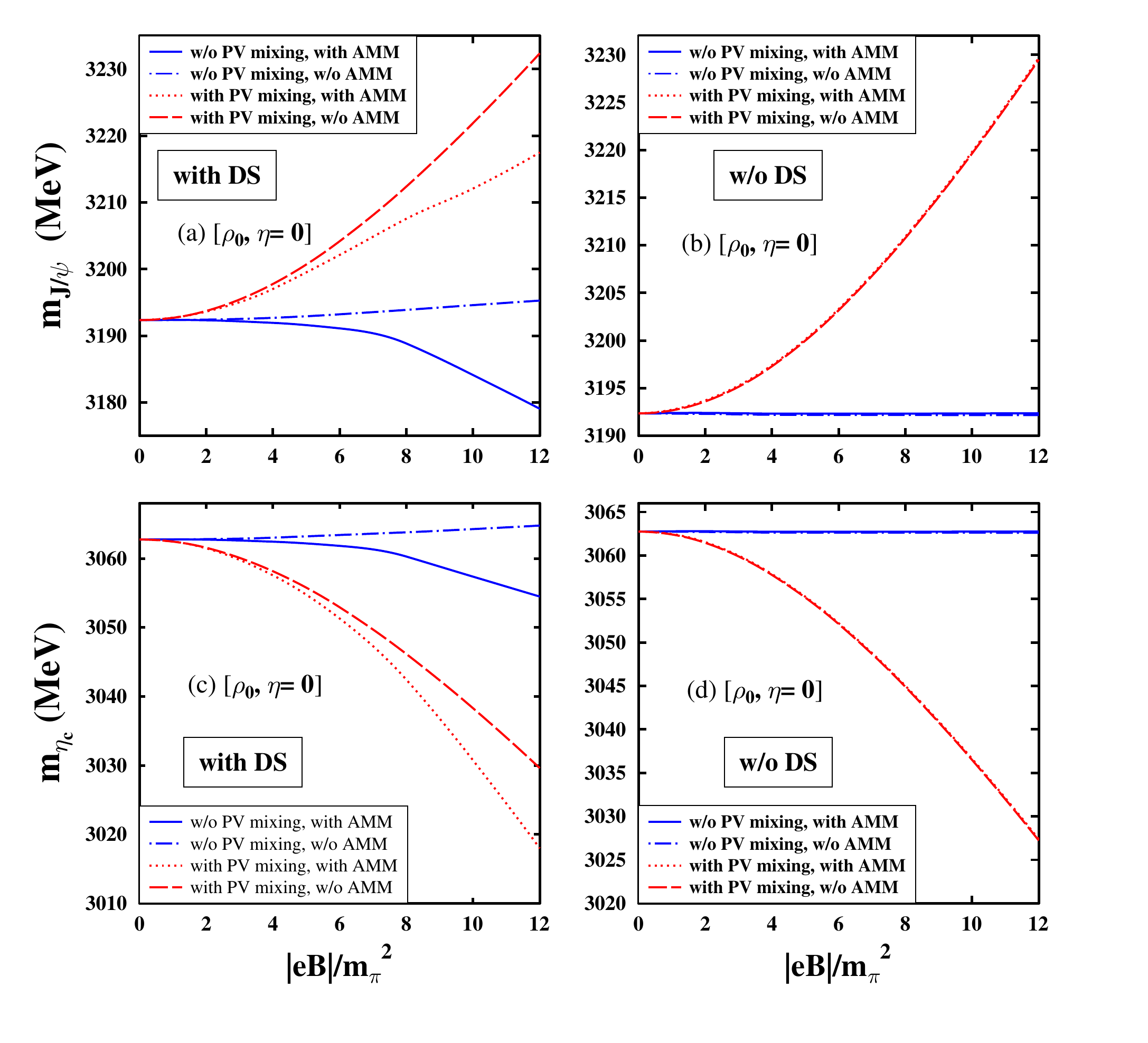}\hfill
    \vspace{-0.8cm}
    \caption{(Color online) Masses (in MeV) are plotted as functions of $|eB|/m_{\pi}^2$ for $J/\psi$ [(a) and (b)], and $\eta_c$ [(c) and (d)] at $\rho_B = \rho_0$, $\eta=0$. The contributions of the magnetized Dirac sea [(a), (c)] to the masses are shown in addition to the effects of the protons' Landau energy levels and the AMMs of the nucleons in magnetized nuclear matter. This is compared to the case when DS effect is absent [(b) and (d)]. The PV mixing effects between $(J/\psi^{||}-\eta_c)$ are considered, accounting for (and not) the effects of  magnetized DS and the AMMs of the nucleons.}
    \label{fig:5}
\end{figure}
\begin{figure}
    \includegraphics[width=1.0\textwidth]{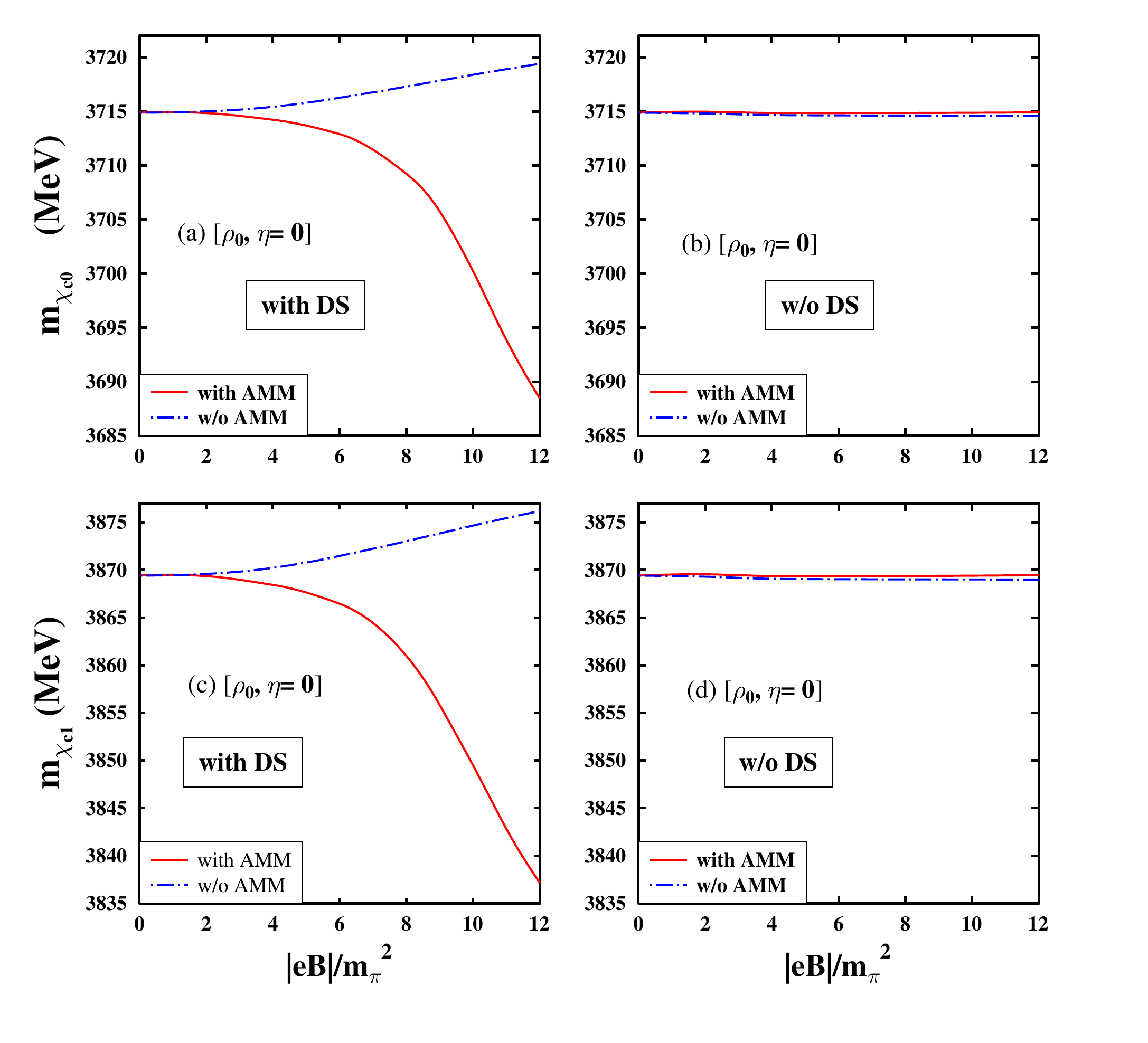}\hfill
    \vspace{-0.8cm}
    \caption{(Color online) Masses (in MeV) are plotted as functions of $|eB|/m_{\pi}^2$ for $\chi_{c0}$ [(a) and (b)], and $\chi_{c1}$ [(c) and (d)] at $\rho_B = \rho_0$, $\eta=0$. The contributions of the magnetized Dirac sea [(a) and (c)] to the masses are shown in addition to the effects of the protons' Landau energy levels and the AMMs of the nucleons in magnetized nuclear matter. This is compared to the case when DS effect is absent [(b) and (d)]. The effects of nucleons' AMMs are taken into account and compared to the case when it is not.}
    \label{fig:6}
\end{figure}
\begin{figure}
    \includegraphics[width=1.0\textwidth]{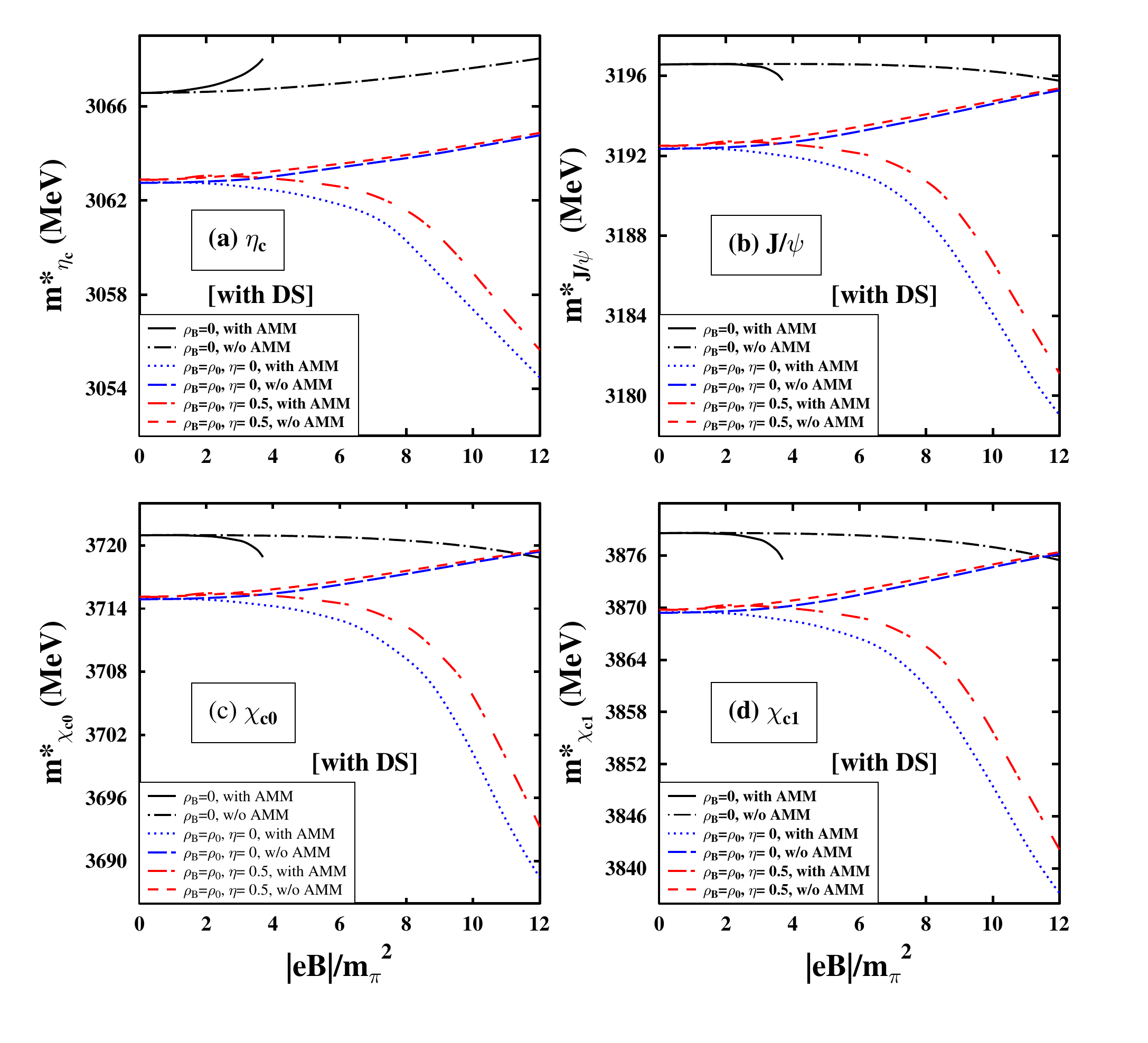}\hfill
    \vspace{-0.8cm}
    \caption{(Color online) Masses of (a) $\eta_c$, (b) $J/\psi$, (c) $\chi_{c0}$ and (d) $\chi_{c1}$ states of charmonium (in MeV) are plotted as functions of $|eB|/m_{\pi}^2$, at $\rho_B=0,\ \rho_0$ for symmetric as well as asymmetric nuclear matter $\eta=0,\ 0.5$. The contributions of Dirac sea (DS) to the masses are studied along with the effects of the Landau energy levels of protons and AMMs of the nucleons in the magnetized nuclear matter. At $\rho_B=0,$ with no Landau quantization of protons only DS effect is there (apart from the PV mixing for the $1S$ wave states in fig.\ref{fig:4}). Masses are shown by considering (and not) the effects of the anomalous magnetic moments of nucleons at finite $|eB|$.}
    \label{fig:7}
\end{figure}

In this subsection, the results for the in-medium masses of the lowest $S$-wave: $J/\psi$ ($^3S_1$), $\eta_c$ ($^1S_0$) and $P$-wave: $\chi_{c0}$ ($^3P_0$), $\chi_{c1}$ ($^3P_1$), states of charmonium are discussed in the presence of magnetized asymmetric nuclear matter, accounting for the effects of the magnetized Dirac sea (denoted as DS). In the sum rule approach, masses are obtained by calculating the moments ($M^i_n$) for all the four meson currents: vector ($^3S_1$), pseudoscalar ($^1S_0$), scalar ($^3P_0$) and axial-vector ($^3P_1$). The moments $M^i_n$, are given in terms of the perturbative Wilson coefficients and the non-perturbative gluon condensates of QCD, as given by equation (10). Wilson coefficients are calculated for different $J^{PC}$ quantum numbers of the currents and are independent of the medium effects \cite{reinders, klingl, song}. The mass formula in the sum rule framework [equation (9)], depends on the running charm quark mass, $m_c(\xi)$, and the running coupling constant, $\alpha_s(\xi)$, which are functions of the renormalization scale $\xi$, given by equations (15) and (16) respectively. The scalar gluon condensate $\langle\frac{\alpha_s}{\pi}G_{\mu\nu}^a G^{a\mu\nu}\rangle$ is connected to the $\phi_b$ term, whereas, the twist-2 gluon condensate $G_2$ to the $\phi_c$ term, which incorporate the medium effects in the mass calculation. The gluon condensates are obtained from the chiral effective model [equations (7) and (8)], in terms of the in-medium scalar fields within the chiral model. In the present investigation, the effects of the magnetized Dirac sea are taken into account through the scalar densities of protons ($\rho^s_p$) and neutrons ($\rho^s_n$). In the magnetized nuclear matter, effects of Landau energy levels of protons and the anomalous magnetic moments of nucleons are also taken into account, which lead to the modifications of the number and scalar densities of nucleons ($\rho_i, \rho^s_i;i=p,n$) \cite{am97, am98, broderik, prakash}. Therefore, within the chiral $SU(3)$ model, the coupled equations of motion in the scalar fields are solved accounting for the protons Landau energy levels and the Dirac sea contribution at finite magnetic field, for the given values of baryon density, $\rho_B$, isospin asymmetry parameter, $\eta=(\rho_n-\rho_p)/2\rho_B$,
($\rho_p$ and $\rho_n$ are the number densities of proton and neutron, respectively) and magnetic field, $|eB|$ (in units of $m_{\pi}^2$). Effects of the magnetized Dirac sea lead to the appreciable changes in the values of the scalar fields within the chiral effective model. In figure (\ref{fig:0}), the values of the light quark condensates $(-\langle \bar{q}q\rangle)^{1/3};\ q=u,d$ (in units of MeV) increase with magnetic field, at $\rho_B=0$, implying magnetic catalysis, with and without the AMMs of the Dirac sea of nucleons. There is observed to be a sharp increase for non zero AMMs of nucleons, as compared to the case of zero AMM. At $\rho_B=\rho_0$, for symmetric as well as asymmetric nuclear matter with $\eta=0$ and $0.5$ respectively, light quark condensates tend to decrease with $eB$ for non zero AMMs of the nucleons, an effect called inverse magnetic catalysis. Its opposite behavior is observed, i.e., increasing values of the light quark condensates with increasing magnetic field for zero AMM, at $\rho_0$ and $\eta=0,0.5$, indicating magnetic catalysis. Thus, the contribution of Dirac sea leads to an inverse magnetic catalysis due to the decreasing values of the light quark condensates with magnetic field within the chiral effective model, at $\rho_0$ for nonzero anomalous magnetic moments of the nucleons. For zero AMM, magnetic catalysis is observed due to the increasing values of the light quark condensates with $|eB|$. It implies that the AMMs of the nucleons play an important role through the magnetized Dirac sea contribution. Along with the light quark condensates, there are observed to be appreciable changes in the magnitudes of the scalar as well as the twist-2 gluon condensates with increasing magnetic field, accounting for the Dirac sea contribution, as shown in figures (\ref{fig:1})-(\ref{fig:2}). In figure \ref{fig:1}, $\langle\frac{\alpha_s}{\pi}G_{\mu\nu}^a G^{a\mu\nu}\rangle^{1/4}$, corresponding to the scalar gluon condensate,  in units of MeV is plotted as a function of magnetic field $|eB|$ in units of $m_{\pi}^2$, at $\rho_B=0,\ \rho_0$ for symmetric and asymmetric nuclear matter, including the effects of the Dirac sea at finite magnetic field, accounting for (and not) the AMMs of the nucleons. The change in the scalar gluon condensate with magnetic field are obtained from the magnetized Dirac sea modified scalar dilaton field $\chi$, as well as of the scalar isoscalar fields $\sigma$ and $\zeta$, from equation (\ref{g1}). The term $(G_2)^{1/4}$ in MeV (with appropriate sign), corresponds to twist-2 gluon condensate is plotted in figure \ref{fig:2}, as functions of $|eB|/m_{\pi}^2$ at $\rho_B=0,\ \rho_0$ for $\eta=0,0.5$, accounting for the DS effects for the $S$- and $P$-wave states of charmonium. Similar pattern are obtained for the $1S$ and $1P$ wave states of bottomonium, as will be described later, for different multiplying factor of $\alpha_s(\xi)$. The running coupling constant $\alpha_s(\xi)$, present in the expression of $G_2$ [equation (7)] are different for $S$ waves with $\xi=1$ \cite{klingl} and $P$ waves with $\xi=2.5$ \cite{song} states of charmonium. In figure (\ref{fig:2}), the variation in the twist-2 gluon condensate $G_2$ is coming from that of the scalar dilaton field $\chi$ and other scalar fields with respect to $|eB|$, following equation (\ref{g2}), accounting for the Dirac sea effects within the chiral model. In the presence of an external magnetic field, the mixing between the longitudinal component of the vector and the pseudoscalar states of charmonium are studied using a phenomenological Lagrangian approach, accounting for the additional effects of Dirac sea to the masses. The effective masses of $J/\psi^{||}$ and $\eta_c$ thus given by equation (\ref{mpv_long}). The nuclear matter saturation density, $\rho_0$ is taken to be $0.15$ fm$^{-3}$ in the present work \cite{papa59}. \\
The value of the renormalization scale, $\xi=1$  is chosen for the $S$-wave \cite{klingl} charmonium states and $\xi=2.5$ \cite{song} for the $P$-wave charmonium states, to study their respective in-medium masses. These choices lead to the $\xi$ dependent coupling constant and charm quark mass,
$\alpha_s=0.21$ and $m_c=1.24$ GeV  for the $1S$-wave states and $\alpha_s=0.1948$ and $m_c=1.22$ GeV for the $1P$-wave states of charmonium. Using the parameters and $\phi_b$ in terms of the scalar gluon condensate, $\langle\frac{\alpha_s}{\pi}G_{\mu\nu}^a G^{a\mu\nu}\rangle$, calculated within the 
chiral effective model, the vacuum masses of $J/\psi$ and
$\eta_c$ are found to be 3196.56 MeV and 3066.57 MeV, respectively. The values of the twist-2 gluon condensate, $G_2$ is zero, and of $\langle\frac{\alpha_s}{\pi}G_{\mu\nu}^a G^{a\mu\nu}\rangle$ is (373.02 MeV)$^4$ at vacuum. The vacuum masses of $\chi_{c0}$ and $\chi_{c1}$ are obtained as 3720.95 MeV and 3878.55 MeV, respectively. The mass shifts of $J/\psi$ at $\rho_B=\rho_0$ and $|eB|=0$ of $-4.21$ MeV calculated in this work can be compared with $-8$ MeV of mass shifts obtained using the leading order QCD formula (similar to second-order Stark effect), in the linear density approximation \cite{ko}.

 For any particular state ($i=$ vector, pseudoscalar, scalar and axial-vector) of heavy quarkonia, the quantity $m_i^*$, in equation (9), in terms of the ratio of two consecutive moments and parameters involving the renormalization scale, $\xi$, can be varied as a function of n. The minimum value of $m_i^*$ corresponds to the physical mass of the state. In our present study, the ratio of two consecutive moments ($M^{i}_{n-1}(\xi)/M^{i}_{n}(\xi)$) in the moment sum rule approach is adopted to obtain the mass of the lowest lying resonance \cite{reinders}. For higher values of n, the effects of the higher lying resonances and continuum can be neglected. However, for large values of n, the incorporation of the higher dimensional operators in the OPE side do not let the first order perturbation theory to hold. To minimize the contributions of the higher dimensional operators, the value of $Q_0^2$ and hence of $\xi$ must be chosen nonzero \cite{reinders85}. Proper choice of $\xi$ sets a range of values for n, which saturates the phenomenological side with the lowest lying resonance \cite{reinders85}. This range is called the stability region in n which changes with $\xi$. For small values of n, breakdown occurs in the stability region due to the contribution of the higher lying resonances. In our calculations, the stability region of n, for the $1S$ and $1P$ states of charmonium and bottomonium are in accordance with the findings of \cite{reinders, reinders85, klingl, song}. In figure \ref{fig:3}, the $m^{*}_{i}$ are plotted as functions of n for the four lowest lying states of charmonium, at $\rho_B=0,\ \rho_0 (\eta=0) $ for $eB=0$. In this figure, the physical mass of the $1S$-waves vector ($J/\psi$) and pseudoscalar ($\eta_c$) states of charmonium are obtained at $n=8$ and $n=9$, respectively for $\xi=1$. For both the $1P$-waves scalar ($\chi_{c0}$) and axial-vector ($\chi_{c1}$) states, it is obtained at $n=9$.

At zero density and finite magnetic field, only Dirac sea effect is there with no protons' Landau level contribution, in the absence of matter part. The masses are calculated by considering the nonzero anomalous magnetic moments (AMMs) of the nucleons and compared to the case when AMM is taken to be zero. At finite density matter, the scalar fields are solved considering the magnetized Dirac sea contribution along with the Landau quantization of protons and non zero anomalous magnetic moments of the nucleons, in presence of magnetic field. The values of the scalar gluon condensate from equation (\ref{g1}), in terms of the scalar dilaton field, $\chi$ and the scalar fields $\sigma$, $\zeta$ (in the limit of finite quark masses), are (in (MeV)$^4$)$, \langle\frac{\alpha_s}{\pi}G_{\mu\nu}^a G^{a\mu\nu}\rangle= (373)^4,\ (372.88)^4,\ (372.48)^4,\ (371.44)^4$ and (370.36)$^4$ at $|eB|/m^2_{\pi}=1,\ 2,\ 3,\ 3.7$ and 3.9, respectively for $\rho_B=0$ and with AMM case. For the case of without AMM, at zero density, values of $\langle\frac{\alpha_s}{\pi}G_{\mu\nu}^a G^{a\mu\nu}\rangle$ are (373.01 MeV)$^4$, (372.93 MeV)$^4$, (372.49 MeV)$^4$ and (371.41 MeV)$^4$ at $|eB|/m^2_{\pi}=2,\ 4,\ 8$ and 12, respectively. In presence of magnetic field, the expectation value of the twist-2 gluon condensate $G_2$ becomes non-zero at $\rho_B=0$ due to the Dirac sea contribution. Thus, the values of $G_2$ in terms of the scalar fields [from equation (\ref{g2})] are given as (57.26 MeV)$^4$ (for $1S$ states), (56.19 MeV)$^4$ (for $1P$ states) at $2 m_{\pi}^2$, and (92.31 MeV)$^4$ (for $1S$), (90.59 MeV)$^4$ (for $1P$) at $3.7 m_{\pi}^2$, with nonzero AMMs of nucleons. In the case of zero AMM, zero density, the values are (35.99 MeV)$^4$ (for $1S$), (35.32 MeV)$^4$ (for $1P$) at $2m_{\pi}^2$, and (74.94 MeV)$^4$ (for $1S$), (73.55 MeV)$^4$ (for $1P$) at $8m_{\pi}^2$. The running coupling constant, $\alpha_s$ is a function of the renormalization scale, $\xi$, thus different choices of $\xi$ for the $S$-wave and $P$-wave states lead to slight variation in $G_2$ values. The values of these condensates are seen to change considerably with the magnetic field, at zero density, accounting for the Dirac sea effect, whereas there is no effect from the Landau quantization of protons. The variation in masses as functions of the magnetic field thus obtained from the modified gluon condensates, as shown in figure (\ref{fig:4}) for $\rho_B=0$, and figures (\ref{fig:5}) ($1S$) to (\ref{fig:6}) ($1P$) at $\rho_0$, $\eta=0$, for the $1S$ ($J/\psi$, $\eta_c$) and $1P$ ($\chi_{c0}$, $\chi_{c1}$) states of charmonium. At finite density matter, $\rho_B=\rho_0$, the contribution from the protons Landau energy levels and anomalous magnetic moments of nucleons are taken into account along with the Dirac sea effects. In figures (\ref{fig:5})-(\ref{fig:6}), masses including the DS effects (plots (a) and (c)) are compared to the case when there is no Dirac sea effect (plots (b) and (d)) for $1S$ and $1P$ wave states of charmonium, taking into account the PV mixing effect of $(J/\psi^{||}-\eta_c)$ for the $1S$ states in (\ref{fig:5}). The parameter $g_{PV}
\equiv g_{ \eta_c J/\psi}$ in the phenomenological Lagrangian (\ref{PVgamma}), is evaluated to be 2.094 from the observed 
radiative decay width of $\Gamma (J/\psi \rightarrow \eta_c \gamma)$
in vacuum, 92.9 keV \cite{pdg}, using equation (\ref{decay_VP}). This effect leads to the rise (drop) in the masses of $J/\psi^{||}$ ($\eta_c$) states with magnetic field as shown in figures (\ref{fig:4}) and (\ref{fig:5}) at $\rho_B=0$ and $\rho_0$ respectively, of amount $0.467(-9.012)$, $10.994(-24.148)$ at $|eB|/m_{\pi}^2=4$ and 8, at $\rho_0$, $\eta=0$ for non zero AMMs of the nucleons, including the Dirac sea effects. The values at $\rho_B=0$ are given as $1.327(-0.99)$, $(3.511)(-2.627)$ at $|eB|/m_{\pi}^2=2,\ 3.5$ for non zero AMMs of the nucleons, incorporating the Dirac sea effects. The differences are taken with respect to their corresponding vacuum mass. The rise and drop in the masses of $J/\psi^{||}$ and $\eta_c$ is identified as a level repulsion between the states with increasing magnetic field. 
%-----------------------------
The level repulsion occurred due to the magnetically induced PV mixing between $J/\psi^{||}$  and $\eta_c$ using an effective hadronic interaction Lagrangian \cite{cho91, am102,cho14}. In the QCD sum rule approach, the magnetically induced PV mixing effects have been incorporated through the current correlator in the phenomenological side and the magnetic field effects through the operator expectation value in the OPE side, leading to the level repulsion between $J/\psi^{||}-\eta_c$ with increasing magnetic field  \cite{cho91, cho14}. The level repulsion from the two approaches of QCDSR and the hadronic effective Lagrangian [in the $2^{nd}$ order of $|eB|$ and leading order in $(\frac{m_V - m_P}{2m_{av}})$ of equation (\ref{mpv_long})], have been observed to be in good agreement in the weak-field region (below $|eB|\sim 0.1\ GeV^2$) with a slight deviation as $|eB|$ increases further \cite{cho91,cho14}. The vacuum masses of $J/\psi$ and $\eta_c$ were found to be $3.092$ GeV and $3.025$ GeV, respectively, for $\langle\frac{\alpha_s}{\pi}G^2\rangle$ of $(0.35\ GeV)^4$, using the Borel transform of sum rule where the Borel curves indicated the level repulsion between $J/\psi$ and $\eta_c$ at $|eB|=0$ and 5$m_{\pi}^2$, accounting for the phenomenologically incorporated mixing effects in the spectral ansatz of QCDSR \cite{cho91,cho14}. In the present work, the vacuum masses for $J/\psi$ and $\eta_c$ are obtained as $3.196$ GeV and $3.067$ GeV by using $\langle\frac{\alpha_s}{\pi}G^2\rangle$ of $(0.37\ GeV)^4$ from the chiral $SU(3)$ model.
%----------------------------
An increase (decrease) in the energy levels of the longitudinal component of $J/\psi$ ($\eta_c$) with magnetic field has been studied due to the spin-magnetic field interaction Hamiltonian within a Cornell potential model \cite{alford}. In an external magnetic field, the quarkonia have a conserved pseudomomentum instead of a conserved center-of-mass momentum. The spin-mixing also lead to the suppression of $J/\psi$ decays to lepton pairs and turn on decays of $\eta_c$ state which should experimentally be realized as a reduction in the dilepton yields of $J/\psi$ and the appearance of a peak at $m_{\eta_c}$ in the dilepton spectrum. In \cite{alford}, an approximate suppression of $11\%$ of the $J/\psi$ decays have been predicted.     
%-------------------------

The in-medium masses of the $J/\psi$, $\eta_c$, $\chi_{c0}$, and $\chi_{c1}$ states of charmonium (decrease) increase with increasing magnetic field at $\rho_0$, due to the (inverse) magnetic catalysis effect (without PV mixing effect of the $1S$ waves), for the case of (nonzero AMMs) zero AMM of the nucleons. In figure (\ref{fig:5}), the masses of $J/\psi$ ($\eta_c$) increase (decrease) with magnetic field, both with and without AMMs of the nucleons in the DS contribution, when the PV mixing effect is considered between ($J/\psi$-$\eta_c$). Although the rate at which mass rises (drops) is modified considerably by the DS effects for non zero AMMs of the nucleons. In comparison to such effects, there are almost no change observed with increasing magnetic field in the absence of Dirac sea effects, due to only the protons' Landau energy levels, and nucleons' AMMs in the magnetized nuclear matter. The observed behavior imply that, an important effect due to the nonzero anomalous magnetic moments of nucleons is coming through the Dirac sea contribution. The mass variation are of similar pattern for the asymmetric nuclear matter ($\eta=0.5$).
%----------
The mass shifts (in MeV) due to the magnetized Dirac sea effects for $J/\psi$ and $\eta_c$ at $\rho_0$ and zero temperature are $-0.41/-3.51/-13.3$ and $-0.31/-2.47/-8.27$ at $|eB|/m_{\pi}^2=4/8/12 $, respectively, for $\eta=0$, which for $\eta=0.5$ are $0.05/-1.72/-11.39$ (for $J/\psi$) and $0.04/-1.32
/-7.24$ (for $\eta_c$). This can be compared with the corresponding mass shifts for $J/\psi$ and $\eta_c$ of $-0.54/-0.84$ and $-0.44/-0.68$ at $|eB|/m_{\pi}^2=5/7 $, for $\eta=0$ and $0.24/0.27$ and $0.20/0.23$ for $\eta=0.5$, respectively, \cite{epja79} in the magnetized nuclear matter, calculated using the QCD sum rule approach and chiral $SU(3)$ model to obtain the medium modified gluon condensates. The shifts are taken with respect to the mass at $\rho_0$ and $|eB|=0$, to show the effects of the magnetic field. The effects of the magnetized Dirac sea on the in-medium masses of $1P$-wave states $\chi_{c0}$ and $\chi_{c1}$ can be inferred from the difference of $m_{\chi_{c0}}(m_{\chi_{c1}})$ between $|eB|/m_{\pi}^2=4$ and $8$ as $-5\ (-7.45)$ MeV, which can be compared with $-2.29\ (-3)$ MeV between $|eB|/m_{\pi}^2=3$ and $7$ in \cite{cpc43}, at $\rho_B=\rho_0$, $T=0$, in the symmetric nuclear matter, using the Borel sum rule.
%------------------
The study of the mass shifts of charmonia due to the medium modifications of the gluon condensates in hot and dense hadronic matter, lead to the downward mass shifts of $J/\psi$ within the QCD perturbative (second-order stark effect) approach \cite{temp85}. The mass shifts calculated using the Borel transformed QCD sum rule lead to approximately twice as large mass shifts for the $\chi_{c1}$ state than for the mass shifts of $J/\psi$ \cite{temp85}, which can be compared with the mass shifts of $-9.14$ MeV for $\chi_{c1}$ which is nearly twice as much as $-4.21$ MeV for the $J/\psi$ calculated at $\rho_B=\rho_0$, $T=0$ and $|eB|=0$, in the present work. The gluon condensates were calculated using a hadronic resonance gas model as compared to the chiral effective model used in the present study. 
%-----------------

In figure \ref{fig:7}, the in-medium masses of (a) $\eta_c$, (b) $J/\psi$, (c) $\chi_{c0}$ and (d) $\chi_{c1}$ charmonium states are plotted as functions of the magnetic field, at $\rho_B=0,\ \rho_0$ for $\eta=0,\ 0.5$ to show the density effects on the masses, accounting for the effects of Dirac sea. In ref. \cite{radfort}, the charmonia spectra (in MeV) of 2980.3 ($\eta_c$), 3097.36 ($J/\psi$), 3415.7 ($\chi_{c0}$) and 3508.2 ($\chi_{c1}$) were obtained perturbatively, in a potential model. The values of the QCD gluon condensates, $\langle\frac{\alpha_s}{\pi}G_{\mu\nu}^a G^{a\mu\nu}\rangle$ and  $G_2$ (for $1S$ and $1P$ states) at $\rho_0$, $\eta=0$ and with AMM, are given respectively, as: (371.78 MeV)$^4$, $-9.7576\times 10^7$ MeV$^4$ ($1S$), $-9.0513\times 10^7$ MeV$^4$ ($1P$) at $2m_{\pi}^2$; (371.58 MeV)$^4$, $-1.0083\times 10^8$ MeV$^4$ ($1S$), $-9.3531\times 10^7$ MeV$^4$ ($1P$) at $4m_{\pi}^2$; and (369.87 MeV)$^4$, $-1.1937\times 10^8$ MeV$^4$ ($1S$), $-1.1073\times 10^8$ MeV$^4$ ($1P$) at $8m_{\pi}^2$; which for the case of without AMM are: (371.83 MeV)$^4$, $-9.6866\times 10^7$ MeV$^4$ ($1S$), $-8.9855\times 10^7$ MeV$^4$ ($1P$) at $2m_{\pi}^2$; (371.9643 MeV)$^4$, $-9.4449\times 10^7$ MeV$^4$ ($1S$), $-8.7613\times 10^7$ MeV$^4$ ($1P$) at $4m_{\pi}^2$; and (372.5342 MeV)$^4$,  $-8.1640\times 10^7$ MeV$^4$ ($1S$), $-7.5731\times 10^7$ MeV$^4$ ($1P$) at $8m_{\pi}^2$. The condensates in the asymmetric nuclear matter ($\eta=0.5$), at the nuclear matter saturation density ($\rho_0$) are given respectively by: (371.97 MeV)$^4$, $ -9.3679\times 10^7$ MeV$^4$ ($1S$), $-8.6899\times 10^7$ MeV$^4$ ($1P$) at $2m_{\pi}^2$; (371.89 MeV)$^4$, $-9.5119\times 10^7$ MeV$^4$ ($1S$), $-8.8234\times 10^7$ MeV$^4$ ($1P$) at $4m_{\pi}^2$; and (370.91 MeV)$^4$, $ -1.0893\times 10^8$ MeV$^4$ ($1S$), $-1.0104\times 10^8$ MeV$^4$ ($1P$) at $8m_{\pi}^2$, for the case of nonzero AMM. The values in case of without AMM are: (371.92 MeV)$^4$, $-9.4527\times 10^7$ MeV$^4$ ($1S$), $-8.7685\times 10^7$ MeV$^4$ ($1P$) at $2m_{\pi}^2$; (372.08 MeV)$^4$, $-9.1366\times 10^7$ MeV$^4$ ($1S$), $-8.4753\times 10^7$ MeV$^4$ ($1P$) at $4m_{\pi}^2$; and (372.61 MeV)$^4$, $-7.8478\times 10^7$ MeV$^4$ ($1S$), $-7.2798\times 10^7$ MeV$^4$ ($1P$) at $8m_{\pi}^2$, for $\langle\frac{\alpha_s}{\pi}G_{\mu\nu}^a G^{a\mu\nu}\rangle$ and  $G_2$ of $1S$ and $1P$ states, respectively. The masses plotted incorporating the effects of magnetized Dirac sea, are denoted as "with DS, with AMM" when the anomalous magnetic moments of the nucleons are considered and "with DS, w/o AMM" for zero AMM. The different behavior of the pseudoscalar meson mass with magnetic field in the absence of nuclear matter, can be attributed to the variation in the Wilson coefficients for different channels.    
\subsection{Bottomonium states}
\begin{figure}[h!]
    \includegraphics[width=1.0\textwidth]{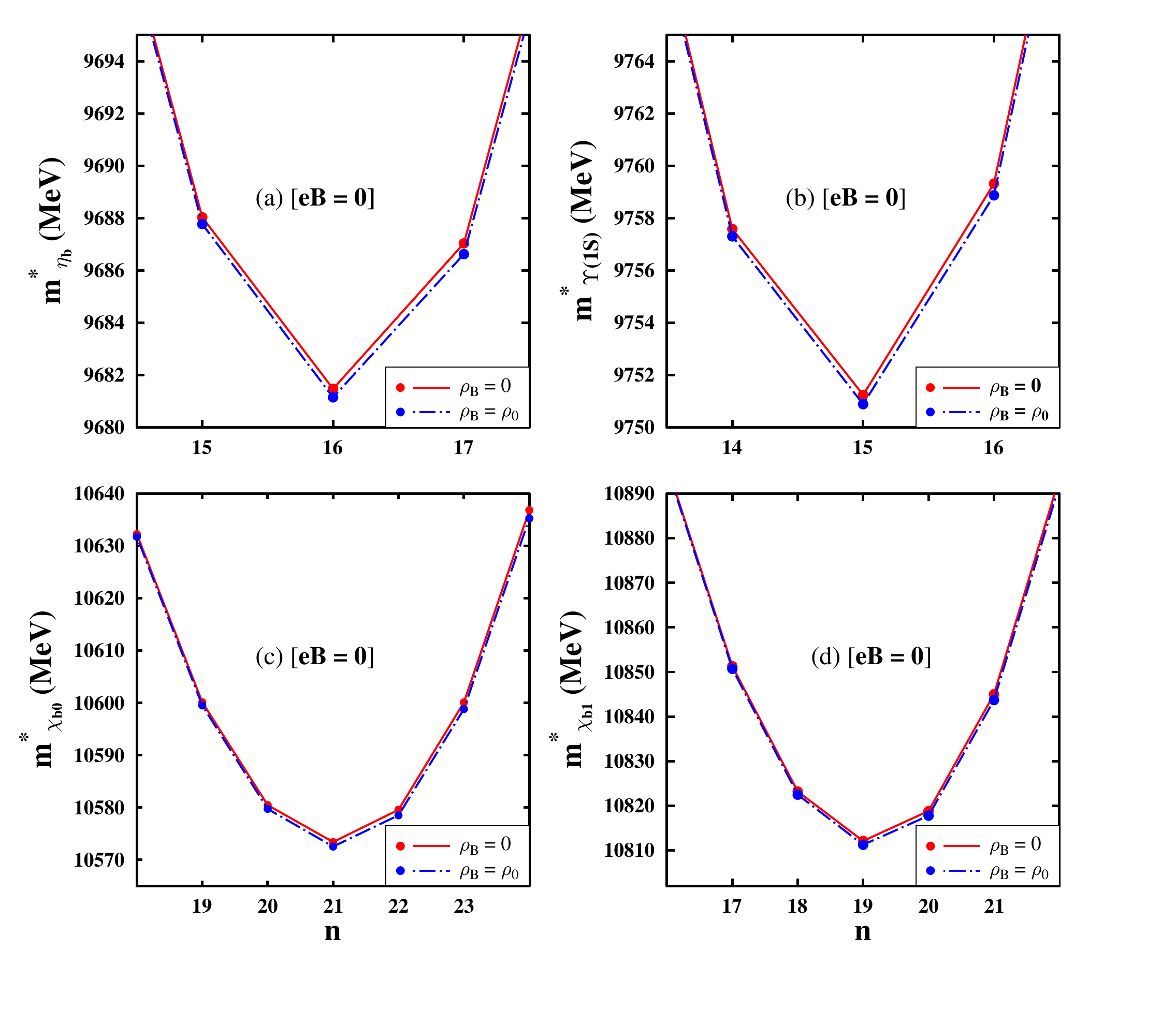}\hfill
    \vspace{-0.8cm}
    \caption{(Color online) $m^*$ (in MeV) are plotted as functions of n, for (a) $\eta_b$, (b) $\Upsilon(1S)$, (c) $\chi_{b0}$ and (d) $\chi_{b1}$ states of bottomonium, at $\rho_B = 0,\ \rho_0$ for symmetric nuclear matter ($\eta=0$), at zero magnetic field $|eB|=0$. The minimum value of $m^{*}$ as a function of n corresponds to the physical mass of that particular state.}
    \label{fig:8}
\end{figure}
\begin{figure}
    \includegraphics[width=1.0\textwidth]{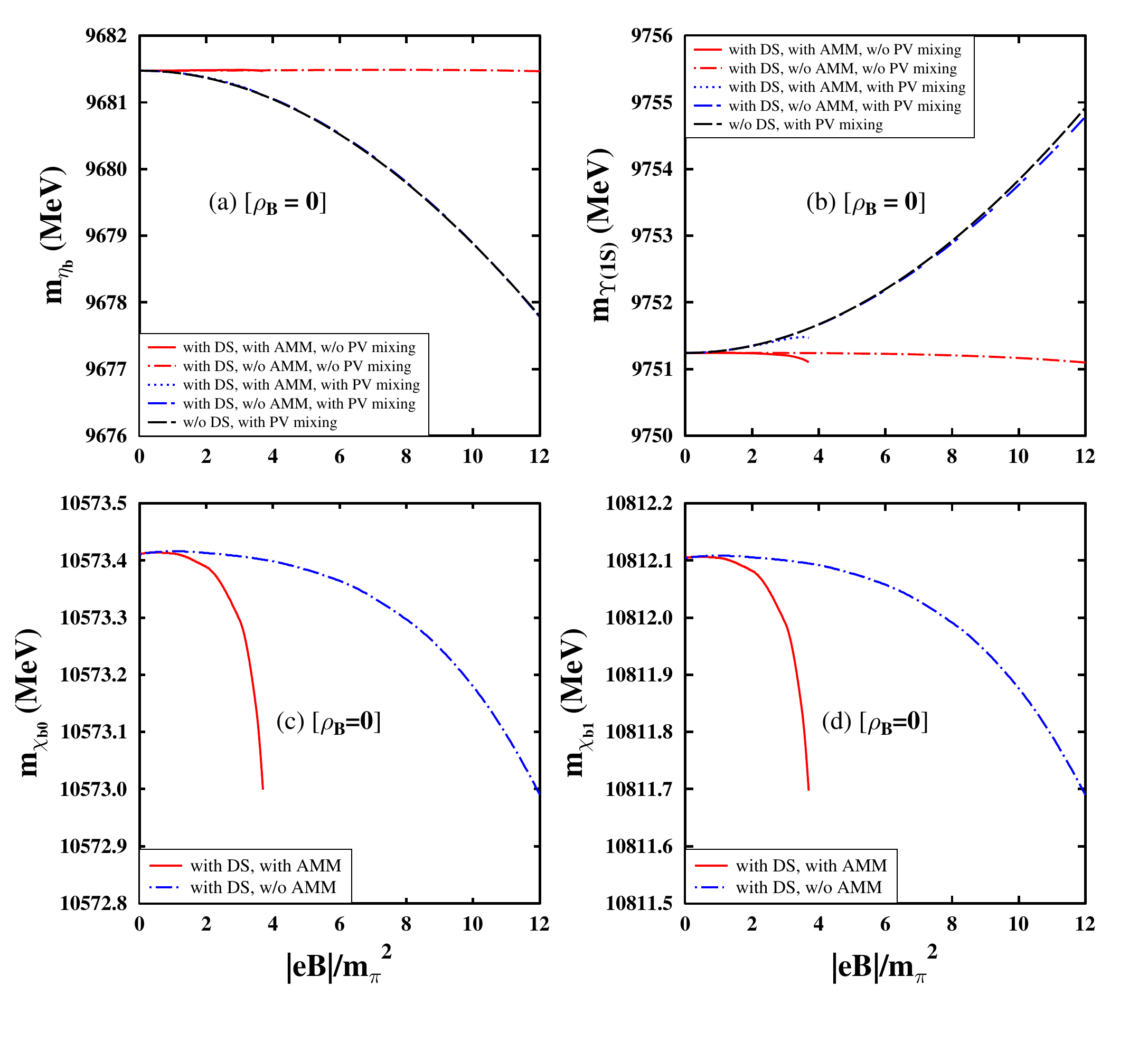}\hfill
    \vspace{-0.8cm}
    \caption{(Color online) Masses (in MeV) are plotted as functions of $|eB|/m_{\pi}^2$, for (a) $\eta_b$, (b) $\Upsilon(1S)$, (c) $\chi_{b0}$ and (d) $\chi_{b1}$ at $\rho_B = 0$. The effects of the magnetized Dirac sea are shown in the masses, accounting for (and not) the anomalous magnetic moments of the Dirac sea of nucleons. The PV mixing effects between $(\Upsilon^{||}(1S)-\eta_b)$ are considered, taking into account the DS effects and compared to the case when it is not included.}
   \label{fig:9}
   \end{figure}
\begin{figure}
\includegraphics[width=1.0\textwidth]{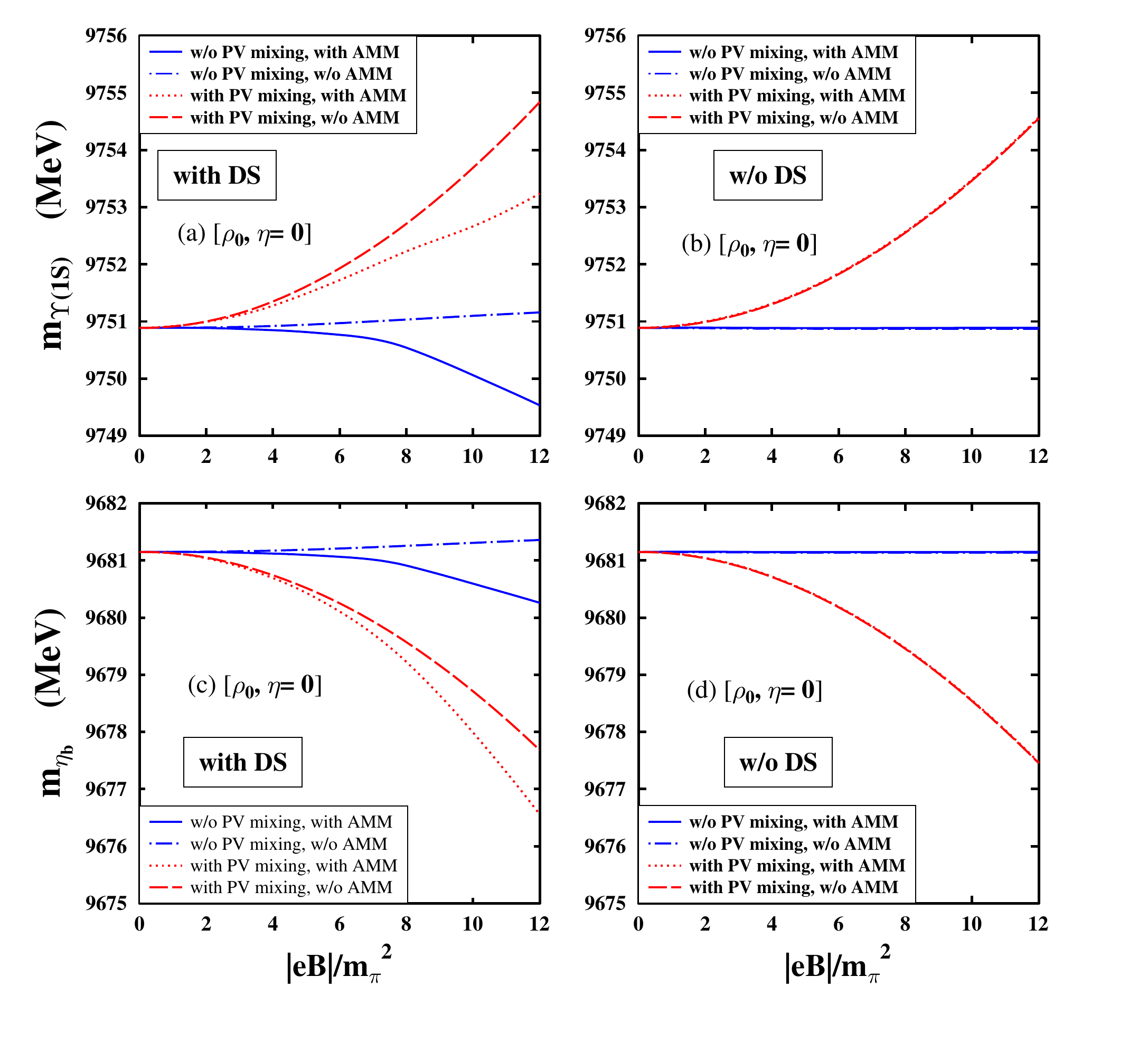} \hfill
\vspace{-1.5cm}
\caption{(Color online) Masses (in MeV) are plotted as functions of $|eB|/m_{\pi}^2$ for $\Upsilon(1S)$ [(a) and (b)], and $\eta_b$ [(c) and (d)] at $\rho_B = \rho_0$, $\eta=0$. The contributions of the magnetized Dirac sea [(a), (c)] to the masses are shown in addition to the effects of the protons' Landau energy levels and the AMMs of the nucleons in magnetized nuclear matter. This is compared to the case when DS effect is absent [(b) and (d)]. The PV mixing effects between $(\Upsilon^{||}(1S)-\eta_b)$ are considered, accounting for (and not) the effects of  magnetized DS and the AMMs of the nucleons.}
   \label{fig:10}
\end{figure}
\begin{figure}
    \includegraphics[width=1.0\textwidth]{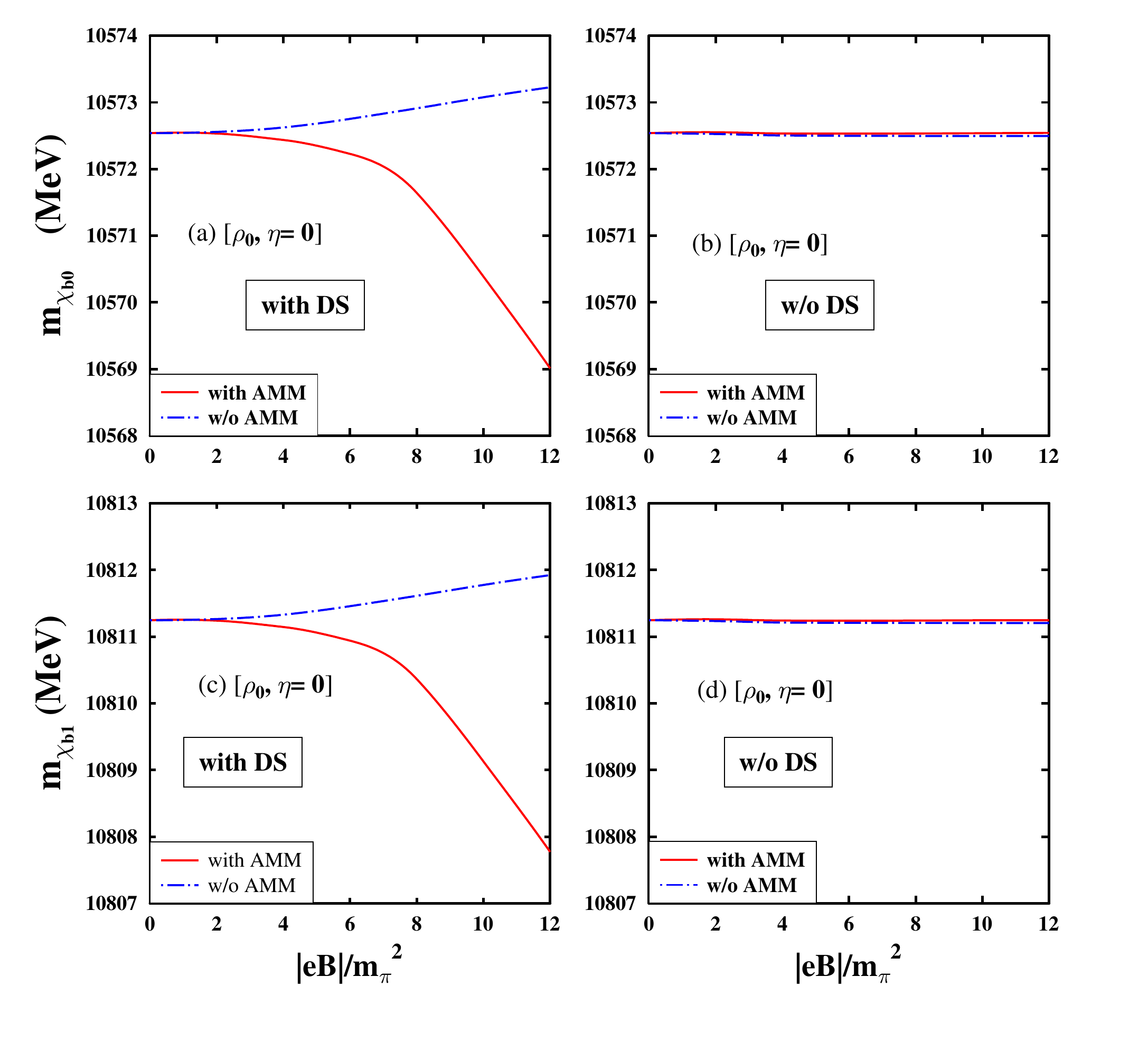}\hfill
    \vspace{-0.8cm}
    \caption{(Color online) Masses (in MeV) are plotted as functions of $|eB|/m_{\pi}^2$ for $\chi_{b0}$ [(a) and (b)], and $\chi_{b1}$ [(c) and (d)] at $\rho_B = \rho_0$, $\eta=0$. The contributions of the magnetized Dirac sea [(a) and (c)] to the masses are shown in addition to the effects of the protons' Landau energy levels and the AMMs of the nucleons in magnetized nuclear matter. This is compared to the case when DS effect is absent [(b) and (d)]. The effects of nucleons' AMMs are taken into account and compared to the case when it is not.}
    \label{fig:11}
\end{figure}
\begin{figure}
    \includegraphics[width=1.0\textwidth]{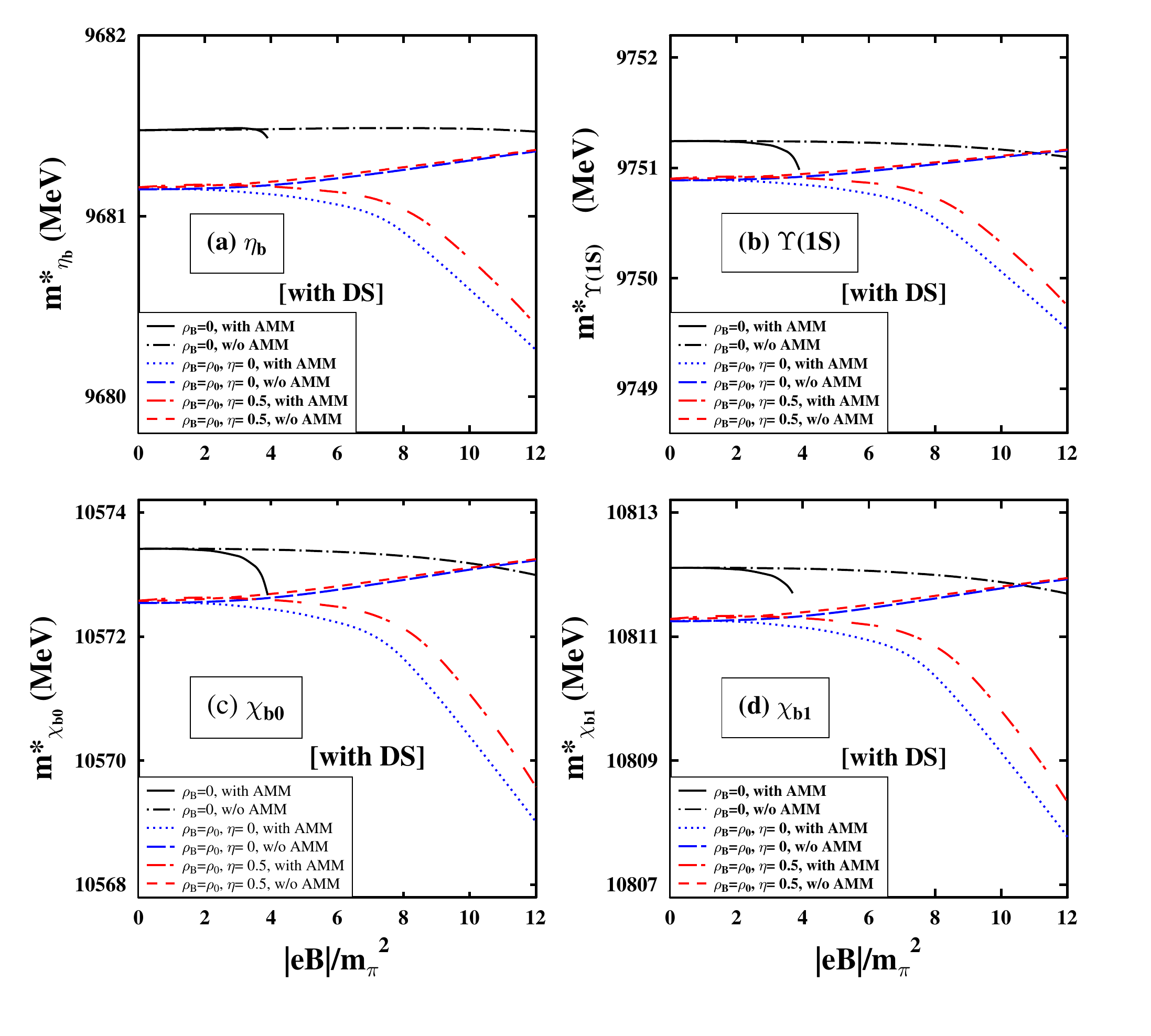}\hfill
    \vspace{-0.8cm}
    \caption{(Color online) Masses of (a) $\eta_b$, (b) $\Upsilon(1S)$, (c) $\chi_{b0}$ and (d) $\chi_{b1}$ states of bottomonium (in MeV) are plotted as functions of $|eB|/m_{\pi}^2$, at $\rho_B=0,\ \rho_0$ for symmetric as well as asymmetric nuclear matter $\eta=0,\ 0.5$. The contributions of Dirac sea (DS) to the masses are studied along with the effects of the Landau energy levels of protons and AMMs of the nucleons in the magnetized nuclear matter. At $\rho_B=0,$ with no Landau quantization of protons only DS effect is there (apart from the PV mixing for the $1S$ wave states in fig.\ref{fig:9}). Masses are shown by considering (and not) the effects of the anomalous magnetic moments of nucleons at finite magnetic field.}
    \label{fig:12}
\end{figure}
In this section, the results for the in-medium masses of the bottomonium ground states, namely the $1S$-wave: $\Upsilon(1S)$ ($^3S_1$), $\eta_b$ ($^1S_0$) and $1P$-wave: $\chi_{b0}$ ($^3P_0$), $\chi_{b1}$ ($^3P_1$), are illustrated in the magnetic asymmetric nuclear matter with the additional contribution from the magnetized Dirac sea. The masses are calculated using the QCD sum rule approach in a similar fashion as described above for the corresponding $1S$ and $1P$ waves states of charmonium. The running bottom quark mass, $m_b(\xi)$ and the running coupling constant, $\alpha_s(\xi)$, as given by equations (15) and (16) respectively, are different from the charmonium states, as the number of current quark flavors is $n_f=5$ for bottom quark, which is $n_f=4$ for the charm quark, also the $\xi$-independent values of the quark mass and couplings (in equations (15) and (16), respectively) are different in the two sectors of charm and bottom quarks. The scalar gluon condensate, $\langle\frac{\alpha_s}{\pi}G_{\mu\nu}^a G^{a\mu\nu}\rangle$ through the $\phi_b$ term and the twist-2 gluon condensate, $G_2$ in $\phi_c$ term, incorporate the effects of density, magnetic fields and isospin asymmetry of the nuclear medium on the bottomonium masses. In the present investigation, we have studied the additional contribution of magnetized Dirac sea on the bottomonium masses. The scalar densities of protons ($\rho^s_p$) and neutrons ($\rho^s_n$) have contributions from the Dirac sea, in addition to the Landau level contributions of protons in the Fermi sea of nucleons. The values of the scalar gluon condensate, $\langle\frac{\alpha_s}{\pi}G_{\mu\nu}^a G^{a\mu\nu}\rangle$ remain same (as can be inferred from equation (\ref{g1})) as it is already discussed, for $\rho_B=0$ as well as $\rho_0$, and for different values of magnetic fields. However, the values of twist-2 gluon condensates, from equation (\ref{g2}), depend on $\alpha_s$, which lead to different values for $G_2$ of $1S$ and $1P$ states of bottomonium. For e.g., at $\rho_B=0$, for non zero AMMs of nucleons, the values of $G_2$ are (51.83 MeV)$^4$ (for $1S$), and (51.23 MeV)$^4$ (for $1P$) at $|eB|=2m_{\pi}^2$, which are different to that in the charmonium sector [(57.26 MeV)$^4$ (for $1S$), and (56.19 MeV)$^4$ (for $1P$) at $|eB|=2m_{\pi}^2$], but the variation in the values with increasing magnetic field are found to be similar in both sectors. Therefore, using the values of $\phi_b$ and $\phi_c$, the in-medium masses of the bottomonium ground states are calculated using equation (9). In presence of an external magnetic field, the spin-magnetic field interaction are studied for the $1S$ states of bottomonium using a Hamiltonian approach \cite{alford}. The effective masses of $\Upsilon^{||}(1S)$ and $\eta_b$, taking into account the spin-mixing effect, are computed using equation (\ref{hm}). \\ Similar to the charmonium sector, the value of the renormalization scale, $\xi=1$  is chosen for the $S$-wave states and $\xi=2.5$ is taken for the $P$-wave states to study their respective in-medium masses. These choices lead to the $\xi$-dependent running coupling constant and running bottom quark mass,
$\alpha_s=0.1411$ and $m_b=4.18$ GeV  for the $1S$ states and $\alpha_s=0.1346$ and $m_b=4.13$ GeV for the $1P$ states of bottomonium, respectively.
From these parameters and with $\phi_b$ calculated within the 
chiral effective model, the vacuum masses (in MeV) of  $\Upsilon(1S)$ and $\eta_b$ are obtained to be 9751.24 and  9681.47 respectively. The vacuum masses (in MeV) of $\chi_{b0}$ and $\chi_{b1}$ are obtained as 10573.41 and 10812.1, respectively.
%------------
The vacuum masses of the lowest lying $\bar{b}b$ states thus obtained can be compared with the position (in GeV) of the lowest peak in the bottomonia spectral function at zero temperature, at $9.56$ ($\Upsilon$), 9.51 ($\eta_b$), 10.15 ($\chi_{b0}$) and 10.42 ($\chi_{b1}$), obtained using the QCD sum rule with maximum entropy method \cite{temp897}. The peak positions are at higher energies than the vacuum masses of these states. However, it is due to the contribution of the excited states along with the ground state in the spectral density function \cite{temp897}, which can also be the case in "pole+continuum" assumption of the usual sum rules where the mass of the ground state are obtained at higher values of energy than the actual vacuum mass of the state \cite{temp897}.   
%-----------
In figure (\ref{fig:8}), $m^{*}_{i}$ for all four bottomonium ground states (a) $\eta_b$, (b) $\Upsilon(1S)$, (c) $\chi_{b0}$ and (d) $\chi_{b1}$, are plotted as functions of the order of the moment, n. The value of n, corresponding to the minimum point gives the physical mass of the associated state, which is for $\eta_b$, $\Upsilon(1S)$, $\chi_{b0}$ and $\chi_{b1}$ is at $n=16,\ 15,\ 21$ and $19$, respectively at $\rho_B=0,\ \rho_0$, $\eta=0$ and for $|eB|=0$. The masses are plotted incorporating the effects of magnetized Dirac sea (DS), are denoted as "with DS, with AMM" when the anomalous magnetic moments of the nucleons are considered and "with DS, w/o AMM" when AMM is taken to be zero. In order to see the importance of the Dirac sea effect on the bottomonium masses, figures (\ref{fig:9}) to (\ref{fig:11}) illustrate the variation of masses with magnetic field. The two situations are compared, when the Dirac sea contribution is there [plots (a) and (c)] along with the protons Landau quantization and when it is absent ("w/o DS") [plots (b) and (d)] in figures (\ref{fig:10}) to (\ref{fig:11}) at finite density. At zero density, only Dirac sea effect is there on the masses in presence of an external magnetic field. The masses of the four lowest lying states of bottomonium have appreciable modifications with increasing magnetic field, when including the DS effects as compared to the no DS effect. In figure (\ref{fig:9}), the masses are plotted as functions of $|eB|/m_{\pi}^2$ for zero baryon density, $\rho_B=0$, accounting for the Dirac sea contribution with the additional spin-magnetic field interaction for the $1S$ states of bottomonium ($\Upsilon(1S)-\eta_b$). The masses are plotted up to $eB = 3.9m_{\pi}^2$ at $\rho_B=0$, for non zero anomalous magnetic moments of the nucleons, as the solutions of the scalar fields including the DS effects are obtained till this point in our current study. It can be attributed to the dominating contribution of the AMMs of Dirac sea of nucleons through $\rho^s_{p,n}$, as the magnetic field increases. Besides the absence of the magnetized Fermi sea part through $\rho_{p,n}$ and $\rho^s_{p,n}$, in the coupled equations of motion of the scalar fields at $\rho_B=0$, lead to unstable solutions for the scalar fields after a certain value of $|eB|$ for non zero AMMs of the nucleons. Thus, the effects of anomalous magnetic moment on the bottomonium states are observed to be quite significant at zero as well as at finite density matter, through the magnetic field modified propagators for the Dirac sea of nucleons. The different behavior of the pseudoscalar meson mass with magnetic field in the absence of nuclear matter, can be attributed to the distinct values of the Wilson coefficients for different meson current channels. In figures (\ref{fig:10}) to (\ref{fig:11}), masses are plotted with magnetic field variation, at $\rho_B=\rho_0$ for symmetric nuclear matter. Similar behavior in masses with slightly different values are obtained in the asymmetric matter for $\eta=0.5$.
%-----------------
The mass shifts of $1S$ and $1P$ wave states of bottomonium, accounting for the Dirac sea effects in the magnetized nuclear matter, at $\rho_0$, $\eta=0$, are (in MeV) $-0.36/-0.56$ (for $\eta_b$), $-0.39/-0.701$ (for $\Upsilon(1S)$), $-0.975/-1.77$ (for $\chi_{b0}$) and $-0.96/-1.74$ (for $\chi_{b1}$) at $|eB|=4/8\ m_{\pi}^2$. The change in $m_{j};\ j=\eta_b, \Upsilon(1S),\chi_{b0},\chi_{b1}$ between $|eB|/m_{\pi}^2=4$ and $8$ of $-0.21$, $-0.31$, $-0.79$ and $-0.78$ MeV can be compared with the in-medium mass change of $\eta_b$, $\Upsilon(1S)$, $\chi_{b0}$ and $\chi_{b1}$ as $-0.13$, $-0.2$, $-0.55$ and $-0.56$ between $|eB|/m_{\pi}^2=3$ and $7$ at $T=0$ \cite{cpc43}, by incorporating the magnetic field effects due to the Landau quantization of protons and AMMs of the nucleons. In \cite{cpc43}, the Borel sum rule has been employed to study the in-medium mass of the lowest lying heavy quarkonia in hot magnetized nuclear matter.   
%------------------
The effects of baryon density as well as of magnetic fields are shown to the masses of (a) $\eta_b$, (b) $\Upsilon(1S)$, (c) $\chi_{b0}$ and (d) $\chi_{b1}$, in figure (\ref{fig:12}), as functions of $|eB|/m_{\pi}^2$, at $\rho_B=0, \ \rho_0$ for $\eta=0,0.5$, accounting for the (inverse) magnetic catalysis. The effects of the spin-magnetic field interaction to the longitudinal component of vector meson, $\Upsilon(1S)$ and pseudoscalar meson, $\eta_b$ are studied here, using a Hamiltonian approach \cite{alford}. This lead to a rise (drop) in the masses of $\Upsilon^{||}(1S)$ ($\eta_b$) with increasing magnetic field, as shown in figures (\ref{fig:9}) [b (a)] for $\rho_B=0$ and (\ref{fig:10}) [a,b (c,d)] for $\rho_B=\rho_0$. The mass of the bottom quark is taken to be $m_b=4.7$ GeV in the bottom quark Bohr magneton, $\mu_b=\frac{|e|/3}{2m_b}$ \cite{alford}. 
%----------------------------
The rise (drop) in the $\Upsilon^{||}(1S)(\eta_b)$ mass with increasing magnetic field should manifests in the suppression of dilepton decays of $\Upsilon(1S)$ state and a profuse production of lepton pairs from $\eta_b$ state due to the states' spin-mixing. In \cite{alford}, approximately $2.8\%$ suppression of dilepton decays of $\Upsilon(1S)$ have been predicted which should instead appear as a peak in the dilepton spectrum at the invariant mass of $m_{\eta_b}$. However, due to the given finite resolution of the detector, it may be experimentally difficult to resolve this properties of $1S$ bottomonium states in the dilepton invariant mass spectrum \cite{alford}. The spectra of heavy quarkonia have been investigated using a potential model \cite{radfort}, consisting of a relativistic kinetic energy term, a linear confining potential with scalar and vector relativistic corrections as well as the perturbative one-loop QCD short distance potential. The masses are calculated using a variational technique. The bottomonia mass spectra (in MeV) are given as 9413.7 ($\eta_b$), 9460.69 ($\Upsilon(1S)$), 9861.12 ($\chi_{b0}$) and 9891.33 ($\chi_{b1}$) \cite{radfort}.  
%----------------------

The general pattern in the mass variation are of similar kind in the charmonium and bottomonium sectors with change in the density and as well as in the magnetic fields, accounting for the (inverse) magnetic catalysis effect, with (without) the PV mixing for the $1S$ states. The effects of magnetic field to the in-medium masses of the heavy quarkonia within the magnetized nuclear medium, are observed to be much more prominent when taken through the Dirac sea contribution for zero and finite baryon density. The magnetic field effects, as seen from the figures, are almost negligible (without PV mixing of $S$ waves) when Dirac sea contribution is not taken into account. The in-medium masses accounting for the PV mixing effect in the $1S$-wave states of $\bar{c}c$ and $\bar{b}b$ mesons, get modified considerably at finite density matter due to AMMs of the nucleons through the Dirac sea. Incorporation of nonzero AMMs of the nucleons through the Dirac sea contribution, lead to significant changes to the masses of the heavy quarkonia in terms of the scalar and twist-2 gluon condensates at zero and finite density matter. Thus the effects of an external magnetic field through the Dirac sea along with the Landau energy levels of protons, and non zero AMMs of the nucleons are considered within the chiral $SU(3)$ model in our present work. The results include appreciable mass modifications of the heavy quarkonium states calculated in the QCD sum rule approach obtained through the medium modified scalar and twist-2 gluon condensates in the chiral model. These mass shifts can significantly modify the in-medium decay widths of charmonium (bottomonium) mesons to open charm (bottom) mesons. This may affect the production of the open heavy flavor mesons as well as of the heavy quarkonia, the formation time of heavy quarkonia, etc. in the peripheral ultra relativistic heavy ion collision experiments where the produced magnetic field is estimated to be large.
\section{Summary}
In the summary, the in-medium masses of the $1S$ and $1P$ waves states of charmonium and bottomonium are studied using the QCD sum rule approach, in the magnetized nuclear medium, accounting for the Dirac sea effects. The medium effects are incorporated through the scalar and twist-2 gluon condensates, calculated in terms of the medium modified scalar dilaton field, $\chi$, and other scalar fields ($\sigma$, $\zeta$, $\delta$), within the chiral effective model. The effects of magnetic field come from the Landau energy levels of protons and the non-zero anomalous magnetic moments of the nucleons, in the magnetized nuclear matter. In the current work, the contribution of the magnetized Dirac sea is incorporated through the scalar densities of nucleons within the chiral effective model. Appreciable modifications in the condensates are obtained due to the Dirac sea effect along with the Landau level contribution of protons, in comparison to the case when Dirac sea effect is not considered. At zero density, the contribution of magnetic field is realized only through the magnetized Dirac sea, with no effect from the Landau energy levels of protons. The nonzero anomalous magnetic moments (AMMs) of protons and neutrons have noticeable effects on the in-medium masses, through the magnetized Dirac sea. In-medium masses of the charmonium and bottomonium ground states, accounting for the Dirac sea effects, are observed to decrease with increasing magnetic field, at finite density matter ($\rho_B=\rho_0$), when AMMs are non-zero, but show opposite behavior (increasing mass with $|eB|$) for zero AMMs of the nucleons. The magnetic fields thus have significant contribution on the in-medium properties (masses and hence on the decay widths) of heavy quarkonia due to the effects of (inverse) magnetic catalysis. This may affect the yield of open heavy flavor mesons and heavy quarkonia in the non-central, high energy heavy ion collision experiments. At finite magnetic field, pseudoscalar-vector mesons (PV) mixing between the longitudinal component of vector and the pseudoscalar meson states, including Dirac sea effects, lead to an upward (downward) shifts in the masses of $J/\psi^{||}$ ($\eta_c$) and $\Upsilon^{||}(1S)$ ($\eta_b$). 
This might be observed as a quasi-peak at the masses of $\eta_c$ and $\eta_b$ states as a consequence of the modified dilepton production. The mixing effect can also lead to a change in the formation time of heavy quarkonia, etc. in the non-central heavy ion collision experiments at RHIC, LHC where huge magnetic fields are generated.

\acknowledgements
Amruta Mishra acknowledges financial support from Department of Science and Technology (DST), Government of India (project no. CRG/2018/002226) and Ankit Kumar from University Grants Commission (UGC), Government of India (Ref. 1279/(CSIR-UGC NET June 2017)).

\end{document}